\documentclass[11pt]{scrartcl}
\usepackage[a4paper, total={16cm, 24cm}]{geometry}

\usepackage{multicol}
\usepackage{tocloft}

\usepackage[small]{caption}

\usepackage{algorithm}
\usepackage{algpseudocodex}
\usepackage[table,dvipsnames]{xcolor}

\usepackage{natbib}
\usepackage[pagebackref]{hyperref}
\hypersetup{
	pdfencoding=auto, 
	psdextra,
	colorlinks=true,
	citecolor=green!50!black,
	linkcolor=red!60!black,
	urlcolor=purple!70!black
}

\renewcommand{\epsilon}{\varepsilon}

\usepackage{authblk}

\author[1]{Kaito Fujii}
\author[2]{Yeeseok Oh}
\author[3]{Shinsaku Sakaue}

\affil[1]{Kyoto University}
\affil[2]{The University of Tokyo}
\affil[3]{CyberAgent, NII, RIKEN AIP}

\usepackage{url}
\usepackage{microtype}
\usepackage[utf8]{inputenc}
\usepackage{graphicx}
\usepackage{amsmath}
\usepackage{amsthm}
\usepackage{amssymb}
\usepackage{mathtools}
\usepackage{booktabs}
\usepackage{enumitem}
\usepackage{todonotes}
\usepackage{wrapstuff}
\usepackage{adjustbox}

\usepackage{wrapfig}

\usepackage[nameinlink]{cleveref}
\AddToHook{cmd/appendix/before}{%
	\crefalias{section}{appendix}%
	\crefalias{subsection}{appendix}
}

\usepackage{stmaryrd}
\usepackage{thm-restate}

\newtheoremstyle{sfthm}
	{\topsep}
	{\topsep}
	{\itshape}
	{}
	{\sffamily\bfseries}
	{}
	{.5em}
	{}
\theoremstyle{sfthm}
\newtheorem{theorem}{Theorem}[section]
\newtheorem{lemma}[theorem]{Lemma}
\newtheorem{proposition}[theorem]{Proposition}
\newtheorem{corollary}[theorem]{Corollary}

\newtheoremstyle{sfdef}
{\topsep}
{\topsep}
{}
{}
{\sffamily\bfseries}
{}
{.5em}
{}
\theoremstyle{sfdef}

\newtheorem{example}[theorem]{Example}
\newtheorem{remark}[theorem]{Remark}

\DeclareMathOperator*{\argmin}{\arg\!\min}
\DeclareMathOperator*{\sc}{\mathrm{sc}}

\newcommand{\Exp}{\mathbb{E}}

\title{Stay or Switch? Metric Distortion with Online Candidate Arrivals}

\date{\vspace{-2.5em}}

\begin{document}

\maketitle

\begin{abstract}
	\begin{center}
		\textbf{\textsf{Abstract}} \smallskip
	\end{center}
	We study metric distortion in sequential elections with online candidate arrivals. In each round, a newly arriving challenger competes against the incumbent, and a voting rule must select one of the two based on voters' pairwise preferences. Our goal is to guarantee low metric distortion at every round, relative to the best candidate among all those that have arrived so far. For deterministic rules, we establish an asymptotically optimal bound of $\Theta(\sqrt{t})$ at round $t$, achieved by a threshold rule favoring the incumbent. In contrast, randomization allows constant distortion. We show that persistent randomized dictatorship achieves optimal distortion of $5$ against adaptive adversaries, which observe previous winners, and $3$ against oblivious adversaries, which do not. Finally, we generalize the model along two dimensions, the available information and the action space, and establish distortion guarantees for each setting.
\end{abstract}

\tableofcontents

\clearpage

\section{Introduction}

Back in the Roman Republic, citizens voted on whether to accept a proposed law. They could choose between \emph{uti rogas} (``as you propose'') and \emph{antiquo} (``I retain the old law''). 
Regarding a single policy or issue, proposals and votes to replace the old law were held successively over many years.
Roman legislation on public land, for instance, underwent successive reforms between 133 and 111 BCE. 
This way of collective decision-making has remained common across political systems, including the U.S. for changes to law. When a new bill is proposed, Congress votes on whether to adopt it or retain the status quo.

Beyond legislative decisions, the structure of incumbent versus challenger elections is a highly common phenomenon in modern elections. 
A recent example comes from the 2026 U.S. midterm elections, in which Senator Susan Collins lost her reelection bid after winning five consecutive Senate elections in Maine since 1996. Such examples illustrate how an incumbent may persist through successive challenges before eventually being replaced. 
Motivated by this common structure of sequential voting, we study collective decision-making under online candidate arrivals, where only the current incumbent and a newly arriving challenger compete for selection.
Can this form of incumbent versus challenger elections guarantee the selection of a good representative at every round?

\subsection{Our Contributions}
We initiate the study of metric distortion with online candidate arrivals.\footnote{An independent, concurrent work studies a similar model; see the related-work section.} In each round, a new candidate arrives, and we have to select a winner for the round.

We consider two dimensions of restrictions, based on the action space and the information structure.
First, we consider the action space. As motivated in the introduction, we introduce the \emph{incumbent-challenger} (IC) model, where at each round, we can either elect the incumbent or the challenger. In contrast, with the full candidate space, we can elect candidates that lost in previous rounds.

Next, we consider the information structure. For the full information setting, we assume that all voters' rankings over the candidates that have arrived so far are revealed. In contrast, in the partial information setting, we only observe the preference of voters between the incumbent and the challenger. The settings are fully explained in \Cref{sec:model}.

Based on the two dimensions, our results are summarized in \Cref{tab:summary}. We consider deterministic rules and randomized rules. For randomized rules, we consider two different adversary models, the adaptive adversary and the oblivious adversary, based on whether the adversary can observe previous winners on the go.

\begin{table}[ht]
\centering
\renewcommand{\arraystretch}{1.4}
\begin{tabular}{@{}lll@{}}
\toprule
\textbf{info \(\backslash\) action}
&
\textbf{full candidate}
&
\textbf{incumbent-challenger}
\\
\midrule
\textbf{full}
&
\begin{tabular}[t]{@{}l@{\quad}r@{\,}c@{\,}l@{}}
(D)        & \multicolumn{3}{l}{$\Delta^* = 3$} \\
(R-Adap.)  & $2.1126$ & $\le \Delta^* \le$ & $2.1248$ \\
(R-Obliv.) & $2.1126$ & $\le \Delta^* \le$ & $2.1248$ 
\end{tabular}
&
\begin{tabular}[t]{@{}l@{\quad}r@{\,}c@{\,}l@{}}
(D)        & \multicolumn{3}{l}{$\Delta^* = 2 + \sqrt{5} \approx 4.2361$} \\
(R-Adap.)  & $3$ & $\le \Delta^* \le$ & $ 2 + \sqrt{5}$ \\
(R-Obliv.) & \multicolumn{3}{l}{$\Delta^* = 3$} 
\end{tabular}
\\
\midrule
\textbf{partial}
&
\begin{tabular}[t]{@{}l@{\quad}r@{\,}c@{\,}l@{}}
(D)        & \multicolumn{3}{l}{$\Delta^* = \Theta (\sqrt{T})$}   \\
(R-Adap.)  & \multicolumn{3}{l}{$\Delta^* = 5$} \\
(R-Obliv.) & \multicolumn{3}{l}{$\Delta^* = 3$}
\end{tabular}
&
\begin{tabular}[t]{@{}l@{\quad}r@{\,}c@{\,}l@{}}
(D)        & \multicolumn{3}{l}{$\Delta^* = \Theta (\sqrt{T})$}  \\
(R-Adap.)  & \multicolumn{3}{l}{$\Delta^* = 5$} \\
(R-Obliv.) & \multicolumn{3}{l}{$\Delta^* = 3$} 
\end{tabular}
\\
\bottomrule
\end{tabular}
\caption{Summary of results for metric distortion bounds on various models. $\Delta^*$ denotes the optimal worst-case distortion guarantee. (D) denotes deterministic voting rules, (R-Adap.) denotes randomized voting rules with adaptive adversaries, and (R-Obliv.) denotes randomized voting rules with oblivious adversaries.}
\label{tab:summary}
\end{table}

Our main focus is on the incumbent-challenger model with partial information. We believe it captures a natural class of repeated binary elections. \Cref{sec:partial-info} derives the bounds for this case, corresponding to the bottom row of \Cref{tab:summary}. 
For deterministic rules, we show a rule that favors the incumbent achieves distortion of $O(\sqrt{T})$, and we show this is asymptotically optimal. Turning to randomized rules, we show that persistent randomized dictatorship achieves constant metric distortion of $5$ and $3$ for adaptive and oblivious adversaries, respectively. We further show these bounds are optimal, proving optimality of persistent randomized dictatorship for metric distortion. This gives the bounds in the bottom-right cell.
In addition, we show allowing the full candidate set as the action space does not improve these bounds, yielding the bottom-left cell.

In \Cref{sec:ic-full-info}, we turn to the case of full information, corresponding to the top row of \Cref{tab:summary}.
With the full candidate setting, the situation coincides with static metric distortion, thus having bounds imported from classical results~\citep{anshelevich2015metric,gkatzelis2020optimal,charikar2022randomized,cai2026stable}. This gives the bounds in the top-left cell.
For the IC model, in contrast to the partial information setting, constant metric distortion of $2 + \sqrt{5}$ is achievable by deterministic rules. This implies that persistent randomized dictatorship is no longer optimal for adaptive adversaries. For oblivious adversaries, however, persistent randomized dictatorship is again optimal. The results give the bounds in the top-right cell.

\subsection{Related Work}

Independently and concurrently with our work, \citet{ghasemi2026distortion} investigate online voting with sequential candidate arrivals. Their model corresponds to our full information, incumbent–challenger model.
\citeauthor{ghasemi2026distortion} also derive the $2 + \sqrt{5}$ upper bound for deterministic rules and $3$ upper bound for randomized rules. They show a deterministic lower bound of $\sqrt{13}$, which is optimal for three candidates. For randomized algorithms, they inherited the static lower bound. These results leave both optimal bounds open. Furthermore, the randomized guarantees are established under oblivious adversaries. In our work, we explicitly consider both oblivious and adaptive adversaries and derive different bounds for each adversary.
Moreover, we close the gaps for their specific model of the full information, incumbent-challenger model, proving that $2 + \sqrt{5}$ is optimal for deterministic rules and $3$ is optimal for oblivious randomized rules. We furthermore characterize optimal distortion for different models.
Their work also considers utilitarian distortion in addition to metric distortion.

\paragraph{Distortion in social choice.}
The notion of distortion was first introduced by \citet{procaccia2006distortion}, measuring the loss of social welfare due to observing ordinal preferences instead of the cardinal utilities. Subsequent work studied the achievable distortion of deterministic and randomized voting rules \citep{caragiannis2011voting,boutilier2015optimal,ebadian2022optimized}; see \citet{anshelevich2021survey} for a survey.

Connecting distortion to the spatial model of voting, \citet{anshelevich2015metric} initiated the study of metric distortion, proving a deterministic lower bound of $3$ and an upper bound of $5$. \citet{munagala2019weighted} improved the upper bound to $2 + \sqrt{5}$ using weighted uncovered sets. The optimal deterministic metric distortion of $3$ was subsequently established by \citet{gkatzelis2020optimal}. \citet{kizilkaya2022plurality} proposed an alternative, simple rule of Plurality Veto that also achieves $3$.
\citet{anshelevich2016randomized} turned to randomized rules and showed that random dictatorship achieves metric distortion of $3$. A sequence of subsequent works has narrowed the gap between the best-known randomized lower and upper bounds \citep{charikar2022randomized,charikar2024barrier,cai2026stable}.

\paragraph{Metric distortion with different ballots.}
Literature on metric distortion considered different types of ballots, often of restricted information. A line of work studies weighted tournament rules, or C2 rules, in which only the pairwise vote counts are observed. \citet{munagala2019weighted} proved metric distortion of $2 + \sqrt{5}$ is achievable, and subsequent work improved the bounds \citep{charikar2025tournament,lin2026weighted}. Distortion has also been studied under approval-based preferences \citep{pierczynski2019approval}. \citet{anshelevich2024threshold} considered additional threshold-approval information, and \citet{frank2026approval} considered approvals under plurality polls. Our partial information setting also restricts the ballots, but in a different way where the comparisons are revealed as candidates arrive, so past choices may limit the information and actions available in subsequent rounds.

\paragraph{Incumbent versus challenger structure.}
Sequential elections with an incumbent versus challenger structure have a long history in the economics and political science literature. The incumbent versus challenger structure was studied as the \emph{amendment procedure}, dating back to \citet{black1948rationale} followed by extensive studies \citep{riker1958paradox,farquharson1969theory,miller1977graph,mckelvey1978multistage,banks1985sophisticated}. These works have traditionally studied the model through the lenses of agenda manipulation, strategic voting, and equilibrium analysis. \citet{bredereck2017parliamentary} study the computational complexity of these problems.

Considering the incumbent-challenger structure as an online process, through competitive analysis or regret analysis, seems highly unexplored. We consider a competitive ratio-based approach to the model via metric distortion.

\paragraph{Online social choice.}
\citet{oren2014online} consider an online setting where voters arrive in an online manner and the decision maker must decide whether to permanently add a candidate to the committee or not.
\citet{dong2025maintaining} consider multiwinner voting under both arrivals and departures of candidates. In ordinal, approval, and metric settings, they propose algorithms that maintain proportionality throughout changes of the candidate set. They also prove impossibility results and handle the cases with algorithms with few replacements.

\section{Model}\label{sec:model}
\subsection{Preliminaries}
For a positive integer $t$, let $[t] = \{1, \ldots, t\}$.
An election instance $E=(N,C,\succ)$ consists of a finite set $N=[n]$ of \emph{voters}, a finite set $C$ of \emph{candidates}, and a \emph{preference profile} $\succ=(\succ_i)_{i \in N}$, where each $\succ_i$ is a strict ranking over the candidates $C$.
A \emph{voting rule} $f$ is a function that takes in an election instance $E=(N,C,\succ)$ as input and outputs the winner $f(\succ) \in C$.

\paragraph{Spatial model of voting.}
In the spatial model of voting, the sets $N$ and $C$ are positioned in a common metric space. In other words, there exists a (pseudo-)metric\footnote{Throughout the paper, we use the terms pseudometric and metric interchangeably.} $d$ on $N \cup C$ satisfying $d(x,x)=0$, $d(x,y)=d(y,x)$, and $d(x,y) \le d(x,z) + d(z,y)$ for all $x,y,z \in N \cup C$.
Voters' preferences towards the candidates are determined by the distances to each candidate, favoring candidates closer to them. 
Given a (pseudo-)metric $d$ on $N \cup C$, we say a preference profile $\succ=(\succ_i)_{i \in N}$ is \emph{induced by} $d$ if for each $i \in N$, $x \succ_i y$ implies $d(i,x) \le d(i,y)$. In this case, we also say that $d$ is \emph{consistent} with $\succ$ and write $d \rhd \succ$.

\paragraph{Metric distortion.}
Given a metric $d$ on $N \cup C$, the \emph{social cost} of a candidate $c \in C$ is $\sc_d(c)=\sum_{i \in N} d(i,c)$. When $d$ is clear by context, we omit the subscript $d$.
Then, we can consider the problem of selecting a candidate that minimizes the social cost.
One can think of a voting procedure as minimizing the social cost. However, since election instances take in ordinal ballots instead of the underlying metric information, we cannot optimize the social cost.
\emph{Metric distortion} measures the approximation ratio to the optimal social cost achievable from ordinal information alone.

Formally, given an election instance $\succ$ and a candidate $x \in C$, the \emph{metric distortion} of $x$ on $\succ$ is defined as
$$\max_{d \rhd \succ} \frac{\sum_{i \in N} d(i,x)}{\min_{y \in C}\sum_{i \in N} d(i,y)}=\max_{d \rhd \succ} \frac{\sc(x)}{\min_{y \in C}\sc(y)},$$
where $0/0 = 1$ and $a/0 = + \infty$ for $a > 0$.
For a voting rule $f$, the \emph{metric distortion} of the rule $f$ is defined as 
$$\Delta(f) =\sup_{(N,C,\succ)}\max_{d \rhd \succ} \frac{\sc(f(\succ))}{\min_{y \in C}\sc(y)},$$
that is, the worst-case approximation ratio of the social cost incurred.
For randomized voting rules where $f(\succ)$ is a probability distribution over $C$, the metric distortion is defined as 
$$\Delta(f) =\sup_{(N,C,\succ)}\max_{d \rhd \succ} \frac{\mathbb{E}[\sc(f(\succ))]}{\min_{y \in C}\sc(y)}.$$

One way we could interpret this value is by considering the following adversary model:
\begin{enumerate}
    \item The adversary selects $N$, $C$, a metric $d$ on $N \cup C$, and a preference profile $\succ$ such that $d \rhd \succ$.
    \item The voting rule $f$ selects the winner $f(\succ)=x \in C$.
    \item The society incurs social cost of $\sum_{i \in N}d(i,f(\succ))$.
\end{enumerate}
We develop this perspective into an online model.

\subsection{Metric Distortion with Online Candidate Arrivals}
Elections are held repeatedly with rounds $t=1,2, \ldots, T$. At each round $t$, a new candidate $c_t$ comes in. We call $c_t$ the \emph{challenger} of round $t$.
Denote by $C_t = \{c_1, \ldots, c_t\}$ the candidates that have arrived by round $t$. Denote by $x_t$ the winner of round $t$. We call $x_{t-1}$ the \emph{incumbent} of round $t$. Note that $x_1 = c_1$ is fixed. 

We consider two models of information or ballots attained at round $t \ge 2$. In the \emph{full information} setting, each voter $i \in N$ reports a ranking on $C_t = \{c_1, \ldots, c_t\}$. In the \emph{partial information} setting, each voter $i \in N$ reports a ranking on $\{x_{t-1}, c_t\}$.
Similarly, we consider two models of action based on the candidate set $X_t$ we can select the winner from on round $t$. In the \emph{full candidate} model, $X_t = C_t = \{c_1, \ldots, c_t\}$. In the \emph{incumbent-challenger} (IC) model, $X_t =\{x_{t-1}, c_t\}$.

The combinations give a total of four possible structures of elections. For each combination, we consider the following problem.

\begin{enumerate}
	\item The adversary selects $N$, $c_1$, a metric $d$ on $N \cup C_1$, and the number of rounds $T$.
	\item For $t=2, \ldots, T$:
	\begin{enumerate}
		\item The adversary places $c_t$, extending $d$ to be defined on $N \cup C_{t}$, and extends the previous preference profile $\succ_{t-1}$ to a preference profile $\succ_t$ such that $d \rhd \succ_t$.
		\item The voting rule $f$ selects the winner $x_t \in X_t$, based on information attained so far.
		\item The society incurs social cost of $\sc(x_t)=\sum_{i \in N}d(i,x_t)$. 
	\end{enumerate}
\end{enumerate}
Note that voter-voter distances cannot differ over rounds, nor may distances between any previously placed points change. For each round $t$, we consider its metric distortion with respect to $C_t$, the candidates so far: 
$$\frac{\sc(x_t)}{\min_{y \in C_t}\sc(y)}.$$
Our objective is to guarantee low metric distortion for every round. Hence, for a (deterministic) rule $f$ that outputs $x_1, \ldots, x_T$, the metric distortion of $f$ is defined as the worst-case distortion among all rounds $t=1, \ldots, T$:
$$\Delta(f) =\sup_{(N,C,\succ), T} \max_{t\in [T]}\max_{d \rhd \succ} \frac{\sc(x_t)}{\min_{y \in C_t}\sc(y)}.$$

One direct observation is that, with full information and full candidate space, the metric distortion bounds coincide with classical metric distortions. This is because we can simply run a static voting rule on each round to achieve the upper bound,\footnote{For randomized static rules against adaptive adversaries, explicitly use fresh random coins at each round.} whereas the same lower bound from a static instance can be derived by introducing the candidates in an arbitrary order. This yields the top-left cell of \Cref{tab:summary}.

\paragraph{Adversaries.}
Due to the sequential nature of the problem, we can consider two types of adversaries.
Upon choosing $c_t$ at round $t$, the \emph{adaptive adversary} chooses $c_t$ with the information of previously selected winners $x_1, \ldots, x_{t-1}$. The \emph{oblivious adversary} chooses $c_t$ without observing previous winners.

In the case of deterministic rules, these two adversaries do not make a difference. An oblivious adversary could simply simulate the voting rule $f$ to predict the winners $x_1, \ldots, x_{t-1}$. For randomized rules, however, distortion could differ by adversary models. For oblivious adversaries, the underlying metric is chosen in advance prior to the rule's realization, and hence the metric distortion becomes 
$$\Delta(f) =\sup_{(N,C,\succ), T} \max_{t\in [T]}\max_{d \rhd \succ} \frac{\mathbb{E}[\sc(x_t)]}{\min_{y \in C_t}\sc(y)}.$$

However, for adaptive adversaries, placement of $c_t$ can depend on previous realizations of $x_1, \ldots, x_{t-1}$, which means that an optimal candidate $y^*_t \in \argmin_{y \in C_t} \sc_d(y)$ at round $t$ could vary due to realizations of $x_1, \ldots, x_{t-1}$. That is, $y^*_t$ itself is a random variable. We hence define the metric distortion as 
$$\Delta(f) =\sup_{\mathcal{A}} \max_{t\in [T]}\frac{\mathbb{E}[\sc(x_t)]}{\mathbb{E}[\sc(y^*_t)]},$$
where the supremum is taken over adaptive adversaries $\mathcal{A}$.

\section{Incumbent-Challenger Model with Partial Information}\label{sec:partial-info}
We consider the case of the incumbent-challenger action model together with partial information. We believe this is the most common case found in real-life examples.

\subsection{Deterministic Voting Rules}

A natural voting rule would be applying the majority rule each round. 
For a single round with two candidates, majority achieves distortion of $3$, and this is optimal.
However for our sequential model, the distortion grows linearly in the number of rounds.
We first introduce a lemma, which is a direct corollary of \citet{Kempe2020LPDuality}.
\begin{lemma}\label{lem:path-ineq}
	Let $a_1, \ldots, a_m, a'$ be candidates such that for each $j = 2, \ldots, m$,
	at least half of the voters prefer candidate $a_{j-1}$ over $a_j$. Let $\alpha = \frac{1}{n} \left| \{ i \in N : a' \prec_i a_{m} \}\right| > 0$. Then,
	$$\sc(a_1) \le \left(2m + \frac{2}{\alpha} -3 \right) \sc(a').$$
	In particular, if $\alpha \ge 1/2$, then $\sc(a_1) \le \left(2m + 1 \right) \sc(a')$.
\end{lemma}
\begin{proof}
	For $m=1$, the case follows trivially from Corollary 5.3~\citep{Kempe2020LPDuality}.
	For $m \ge 2$, let $p_j$ ($2 \le j \le m$) be the fraction of voters preferring $a_{j-1}$ over $a_j$. 
	Using the notation of Corollary 5.1~\citep{Kempe2020LPDuality}, we have numbers $\lambda_1 =1, \lambda_2 = \frac{2}{p_2} -1 \le 3, \lambda_j = \frac{2}{p_j} \le 4$. Then $\Lambda$, the largest sum of numbers in $(\lambda_1, \ldots, \lambda_m, 2/\alpha)$ with no two consecutive numbers, satisfies $\Lambda \le 2m + \frac{2}{\alpha} -3$, and hence the result follows.
\end{proof}

\begin{proposition}\label{prop:majority}
	Round-wise majority achieves metric distortion at most $2T-1$ in the partial information IC model, and this is tight.
\end{proposition}
\begin{proof}
	At each round $t$, add a directed edge between $c_t$ and $x_{t-1}$ from the winner of majority to the loser. Then the resulting underlying undirected graph is a tree, since it is connected and has $T-1$ edges with $T$ vertices.
	Note that the incumbent $x_t$ at round $t$ has a directed path to all candidates $\{c_1, \ldots, c_t\} \setminus \{x_t\}$ that have arrived by round $t$.
	This can be shown by the following induction on $t$. If $x_t = x_{t-1}$, there is a directed edge $x_{t-1}c_t$, and other candidates are reachable by the induction assumption. If $x_t = c_t$, there exists a directed edge $c_t x_{t-1}$, and other candidates are reachable through $x_{t-1}$.
	Hence, for each $i \in [t]$, there exists a directed path from $x_t$ to $c_i$ with length at most $t-1$. By \Cref{lem:path-ineq}, we have $\sc (x_t) \le (2t-1) \sc(c_i)$ for all $i \in [t]$, for all $t \in [T]$.

	We now give a matching lower-bound example. Consider the following instance. First place $2M$ voters on a line: $M$ voters at $0$ and $M$ voters at $1$. Place candidates $c_1, \ldots, c_T$ at $0, 2, -2, 4, -4, \ldots$, i.e., place $c_t$ at $t$ if $t$ is even and $1-t$ if $t$ is odd.
	Finally, place an additional voter $w$ whose distance to every voter or candidate on the line is $T$. Note that the resulting metric is a valid (pseudo-)metric: for any line-points $y,z$, $d(y,z) \le 2T = d(w, y) + d(w, z)$ and $d(w,y) = T \le d(w,z) + d(z,y)$.
	
	For each voter, break equal distances in favor of later arriving candidates.
	Now observe that at every round $t$, the challenger $c_t$ wins: $M$ voters at $0$ (if $t$ is odd), or $M$ voters at $1$ (if $t$ is even), together with voter $w$ always prefer $c_t$ over $x_{t-1}$. However, we have $\sc(c_T) = M(2T-1) + T$ and $\sc(c_1) = M + T$. Hence, as $M \to \infty$,
	\begin{equation*}
			\frac{\sc(c_T)}{\sc(c_1)} \to 2T - 1. \qedhere
	\end{equation*}
\end{proof}

Can we obtain better distortion? It turns out by altering the threshold for the number of votes needed for the challenger to win in an incumbent-favorable way, we can decrease the distortion to $O(\sqrt{T})$.
Furthermore, this is asymptotically optimal in $T$.

\begin{algorithm}[t]
	\begin{algorithmic}
		\State \textbf{Input}: An online voting instance of $T$ rounds and a nonincreasing sequence $(\alpha_t)_{t \ge 2}$ satisfying $0 < \alpha_t \le 1/4$
		\State \textbf{Output}: A sequence of winners $x_1,\ldots,x_T$ satisfying $x_t\in\{x_{t-1},c_t\}$ for every $t\ge 2$
		\State $x_1 \gets c_1$
		\For{$t=2,\ldots,T$}
		\State $q_t \gets \frac{1}{n}|\{i \in N : x_{t-1} \succ_i c_t\}|$
		\If{$q_t \le \alpha_t$}
		\State $x_t \gets c_t$
		\Else
		\State $x_t \gets x_{t-1}$
		\EndIf
		\EndFor
		\State \Return $(x_1,\ldots,x_T)$
	\end{algorithmic}
	\caption{$\alpha$-threshold majority}
	\label{alg:higher-threshold-rule}
\end{algorithm}

\begin{theorem}
	Let  $(\alpha_t)_{t \ge 2}$ be a nonincreasing sequence satisfying $0 < \alpha_t \le 1/4$.
	Under \Cref{alg:higher-threshold-rule}, for every $t \ge 2$ and $j \le t$,
	$$ \frac{\sc (x_t)}{\sc (c_j)} \le \frac{2}{\alpha_t} + 8 \sum_{s=2}^t \alpha_s + 1.$$
\end{theorem}
\begin{proof}
	Construct the weighted directed graph as follows. At round $t$, add a directed edge from the candidate chosen $x_t$ to the loser. If $x_t = c_t$, call the edge a switch edge and assign weight $1-q_t$. If $x_t = x_{t-1}$, call the edge a rejection edge and assign weight $q_t$.
	That is, we assign the weights to be the proportion of voters preferring $x_t$ over the loser.
	
	As in \Cref{prop:majority}, the graph is a directed tree where there exists a directed path from $x_t$ to each candidate $c_j$, $j \in [t]$.
	Furthermore, the directed path from $x_t$ to $c_j \ne x_t$ consists of consecutive switch edges with possibly the last edge being a rejection edge. This is because the endpoint of a rejection edge becomes a leaf.
	
	Fix a round $t \ge 2$ and $j \le t$, and consider the directed $x_t$-$c_j$ path. 
	Let the switch edges of the $x_t$-$c_j$ path be $e_1, \ldots, e_\ell$, in path order. We partition the edges into blocks of consecutive edges in the following way:
	For a switch edge $e_p$, let the weight be $1 - \alpha_p'$. Greedily extend each block as far as possible subject to the sum of all $\alpha_p'$ over edges $p$ in the block being at most $1/2$. Let $R$ be the number of blocks. (Let $R=0$ if there are no switch edges.)
	
	Note that with $\ell$ consecutive switch edges with weights $1 - \alpha_1', 1 - \alpha_2', \ldots, 1- \alpha_\ell'$, by the union bound, there are at least $1 - \sum_{p=1}^\ell \alpha_p'$ proportion of voters who support all edges. Hence if $\sum_{p=1}^\ell \alpha_p' \le 1/2$, by transitivity, at least half of the voters prefer the source candidate to the sink candidate. Hence, we can replace each block by a majority edge with weight $\ge 1/2$. This yields an $x_t$-$c_j$ path with $R$ edges with weight $\ge 1/2$ and possibly a final rejection edge with weight $q_j \ge \alpha_j$. Since $j \le t$, we have $\alpha_j \ge \alpha_t$.
	Hence, by \Cref{lem:path-ineq}, we have
	$$\sc(x_t) \le \left(2R + \frac{2}{\alpha_t} - 1 \right) \sc(c_j).$$
	
	Finally, we bound $R$. 
	Every block except the final one is maximal: adding the $\alpha$ of the next edge makes the sum of $\alpha$'s exceed $1/2$. Since each $\alpha$ is at most $1/4$, each non-final block has sum of $\alpha$'s at least $1/4$.
	Hence the sum of $\alpha$'s in the switch edges of the original $x_t$-$c_j$ path is at least $(R-1) \cdot \frac{1}{4}$. Since the sum of $\alpha$'s is bounded by $\sum_{s=2}^t \alpha_s$, we have $R \le 4 \sum_{s=2}^t \alpha_s + 1$. Hence, we attain the bound.
\end{proof}

By setting $\alpha_t = 1/(4\sqrt{T})$ for all $t$, we obtain a metric distortion of $O(\sqrt{T})$.
In fact, we can guarantee a distortion of $O(\sqrt{t})$ at round $t$, even without the knowledge of the horizon $T$.

\begin{corollary}
	\Cref{alg:higher-threshold-rule} with $\alpha_s = 1/(4\sqrt{s})$ achieves metric distortion of $O(\sqrt{t})$ at round $t$ in the partial information IC model.
\end{corollary}
\begin{proof}
	Set $\alpha_s = 1/(4\sqrt{s})$. Then, 
	\begin{align*}
		\frac{2}{\alpha_t} + 8 \sum_{s=2}^t \alpha_s + 1 &= 8 \sqrt{t} + 2 \sum_{s=2}^t \frac{1}{\sqrt{s}} + 1 \\
		&\le 8 \sqrt{t} + 2 \int_{1}^t \frac{1}{\sqrt{x}}dx + 1 \\
		&= 12 \sqrt{t} -3.  \qedhere
	\end{align*}
\end{proof}

Furthermore, the bound is asymptotically tight; deterministic voting rules cannot escape from distortion accumulating at speed of the square root of the number of rounds.

\begin{proposition}
	Every deterministic voting rule has metric distortion $\Omega(\sqrt{T})$ in the partial information IC model. 
\end{proposition}

\begin{proof}
	Fix a deterministic rule. For an integer $k \ge 2$, let $T=k(k-1)+1$.
	Consider $k$ voters and $T$ candidates arriving in order of $c_1,\ldots,c_T$.
	At round $t$, only voter $i_t = 1 + ((t-2) \bmod k)$ prefers $x_{t-1}$ over $c_t$, whereas all other $k-1$ voters prefer $c_t$. We divide into cases according to whether the rule ever selects an incumbent.
	
	\paragraph{Case 1: The rule always chooses the challengers.} 
	Define the voter--candidate distances by $d(i,c_t) = 1+2\left\lfloor\frac{t-1+k-i}{k}\right\rfloor$.
	In case of ties, break them in favor of larger $t$.
	Then, for every $2\le t\le T$,
	$$d(i, c_t) - d(i,c_{t-1}) =
	\begin{cases}
		2, & i= i_t,\\
		0, & i \ne i_t.
	\end{cases}
	$$
	Hence voter $i_t$ prefers $c_{t-1}$ to $c_t$, while every other voter prefers $c_t$ by the specified tie-breaking.
	Define other distances using the shortest-path distance.
	
	Let us verify that the resulting metric satisfies the metric axioms.
	Since we have defined distances on the complete bipartite graph of voters and candidates, it is sufficient to check for $3$-path triangle inequalities: $d(i,c) \le d(i,c') + d(i',c') + d(i',c)$ for all $i, i' \in N$ and $c, c' \in C$.
	For every $t$ and every pair of voters $i, i'$, we have $d(i,c_t)\ge 1$ and $|d(i,c_t) - d(i', c_t)|\le 2$. Hence,
	$d(i,c) - d(i',c) \le 2 \le d(i,c') + d(i', c')$, and the $3$-path triangle inequality holds.
	
	Now let us compute the distortion. Since $T - 1 = k(k-1)$, for every voter $i$, we have $d(i,c_1) = 1$ and $d(i,c_{T}) = 2k -1$. Since $x_{T} = c_{T}$, we have
	$$\frac{\sc(x_{T})}{\sc(c_1)} = \frac{k(2k-1)}{k} = 2k-1 = \Theta(\sqrt{T}).$$

	\paragraph{Case 2: The rule chooses the incumbent at least once.}
	Let $t$ be the first round the rule chooses the incumbent $x_{t-1}$. Note that $x_{t-1} = c_{t-1}$.
	Then, consider the following instance where the voters and the candidates are on the line.
	Place candidate $c_t$ at $2$ and all other candidates $c_s \ne c_t$ at $0$.
	Place voter $i_t$ at $1$ and all other voters $i \ne i_t$ at $2$.
	Breaking ties in a favorable way where all voters except $i_s$ favor candidates that arrive late at round $s$, the specified preference profile above is induced.
	
	Since all candidates except $c_t$ are placed at $0$, and $c_t$ is never a winner, $x_{T}$ is located at $0$. Hence,
	$$\frac{\sc(x_{T})}{\sc(c_t)} = \frac{2(k-1)+1}{1} = 2k-1 = \Theta(\sqrt{T}).$$

	Note that for arbitrary $T$, we obtain the same asymptotic bound by setting $k=\lfloor \sqrt{T} \rfloor$, applying the above argument with $T_0 = k(k-1)+1$, and padding with dummy candidates.
\end{proof}

\begin{remark}
	Enlarging the candidate space to $C_t= \{c_1, \ldots, c_t\}$ does not help; the above lower bound instance still applies for this larger action space by evaluating distortion at the first round $t$ at which the rule selects an old candidate instead of the challenger. Since voting rules for the IC model also work under full candidate space, distortion of $\Theta(\sqrt{T})$ is optimal for both partial information models.
\end{remark}

\subsection{Randomized Voting Rules}
Allowing randomization yielded a separation for static metric distortion. Will this still be the case for the online model? 

We adapt randomized dictatorship to our setting, which gives metric distortion of $3$ in the static setting~\citep{anshelevich2016randomized}.
One way would be selecting a random dictator for each round $t$. However, even allowing full candidate space $C_t$, we cannot necessarily find out the most preferred candidate of the dictator among the candidates that have arrived so far. This is because we only observe the preference between the incumbent $x_{t-1}$ and the challenger $c_t$. Instead, we could select the candidate the dictator prefers among $\{x_{t-1}, c_t\}$. This, however, suffers distortion linear in the number of rounds $T$.

\begin{example}[Linear distortion under roundwise randomized dictatorship]
	Consider placing $n$ voters on a line with $n-1$ voters at $0$ and one voter $i^*$ at $1$. Place $c_1$ at $0$ and $c_t$ at $2$ for $t \ge 2$. Break ties so that $i^*$ prefers $c_2, c_3, \ldots$ over $c_1$. While the incumbent is $c_1$, if $i^*$ is selected as the dictator, then the winner will switch to the challenger, while if the dictator is not $i^*$, the incumbent remains. Consequently, $\Pr(x_T \ne c_1) = 1 - (1 - 1/n)^{T-1}$. Hence, distortion at round $T$ is
	$$\frac{\Exp[\sc(x_T)]}{\sc(c_1)} = 1 + 2(n-1) \left( 1 - \left(1 -\frac{1}{n} \right)^{T-1} \right).$$
	Taking $n \to \infty$ for fixed $T$, we have distortion reaching $2T - 1$.
\end{example}

Instead, we consider keeping the dictator persistent through every round: \Cref{alg:persistent-random-dictatorship}. With the dictator fixed, we can infer the top choice of the dictator through transitivity.
\begin{algorithm}[t]
	\begin{algorithmic}
		\State \textbf{Input}: An online voting instance of $T$ rounds
		\State \textbf{Output}: A sequence of winners $x_1,\ldots,x_T$ satisfying $x_t\in\{x_{t-1},c_t\}$ for every $t\ge 2$
		\State Sample a voter $i^\star\in N$ uniformly at random
		\State $x_1 \gets c_1$
		\For{$t=2,\ldots,T$}
		\If{$c_t \succ_{i^\star} x_{t-1}$}
		\State $x_t \gets c_t$
		\Else
		\State $x_t \gets x_{t-1}$
		\EndIf
		\EndFor
		\State \Return $(x_1,\ldots,x_T)$
	\end{algorithmic}
	\caption{Persistent random dictatorship}
	\label{alg:persistent-random-dictatorship}
\end{algorithm}
A direct observation is that under oblivious adversaries, we obtain the same guarantee as in the static case.

\begin{proposition}
	Under oblivious adversaries, \Cref{alg:persistent-random-dictatorship} achieves metric distortion at most $3$ in the partial information IC model.
\end{proposition}
\begin{proof}
	\Cref{alg:persistent-random-dictatorship} selects the best candidate for the randomized dictator $i^\star$ each round, which achieves metric distortion of at most $3 - \frac{2}{n}$~\citep{anshelevich2016randomized}.    
\end{proof}

The interesting case is under adaptive adversaries. Such an adversary can learn the persistent dictator through rounds of play and yield a worse distortion.
\begin{example}[Learning the persistent dictator]
	Consider the adversary placing $n$ voters on the line with voter $i$ placed at $i$. At round $t$, place $c_t$ at $t$. Let $i^*$ be the persistent dictator. Then, the winners are $(c_1, c_2, \ldots, c_{i^*}, c_{i^*}, \ldots)$, that is, the dictator always selects the challenger until round $i^*$ and then keeps on selecting the incumbent. Hence, after at most $n$ rounds, we can determine who the dictator is.
\end{example}
After learning who the dictator is, the adversary could introduce candidates that maximize distortion.
Surprisingly, however, \Cref{alg:persistent-random-dictatorship} still obtains constant metric distortion of $5$.
From the adversary's perspective, this resembles the \emph{exploration-exploitation tradeoff} in reinforcement learning. The adversary can try to explore who the dictator is, which has the cost of having bounded exploitation to cause bad distortion.

\begin{theorem}
	Under adaptive adversaries, \Cref{alg:persistent-random-dictatorship} achieves metric distortion at most $5$ in the partial information IC model.
\end{theorem}
\begin{proof}
	The case is trivial for $n \le 2$. Assume $n\ge 3$.
	Fix a round $t$ and an adaptive adversary.
	For the case where the dictator is chosen to be $i \in N$, let $x_{i}$ be the candidate selected at round $t$, i.e. $x_{i}$ is ranked top by $i$ among $C_t$. Let $d_{i}$ be the metric at round $t$ chosen by the adversary when the dictator is chosen to be $i$, and let $y_{i}^*$ be an optimal candidate that minimizes the social cost at round $t$ w.r.t. $d_{i}$.
	
	The metric distortion at round $t$ is then
	$$\frac{\mathbb{E}\left[\sc(x_t)\right]}{\mathbb{E}\left[\sc(y_t^*)\right]} = \frac{\frac{1}{n}\sum_{i \in N}\sc_{d_{i}}(x_{i})}{\frac{1}{n}\sum_{i \in N} \sc_{d_{i}}(y_{i}^*)}.$$
	
	First note that we have 
	$$\sum_{i \in N}\sc\nolimits_{d_{i}}(y_{i}^*) = \sum_{i,j \in N}d_{i}(j,y_i^*)= \sum_{i \in N}d_{i}(i,y_{i}^*) + \sum_{i \ne j}d_i(j,y_i^*).$$
	
	On the other hand, since $d_i(j, x_{i}) \le d(j,i)+d_i(i,x_{i}) \le d(j,i)+d_i(i,y_i^*)$, we have 
	$$\sum_{i \in N}\sc\nolimits_{d_{i}}(x_{i}) = \sum_{i, j \in N}d_{i}(j, x_{i}) \le \sum_{i \ne j}d(j,i)+ n \cdot \sum_{i \in N} d_i(i,y_{i}^*).$$
	Note that since voter-voter distances are common across realizations of the random dictator, we can simply write $d(j,i)$ without a subscript. 
	Hence, we have
	$$\sum_{i \ne j}d(j,i)=2 \sum_{i<j}d(j,i).$$
	
	Fix $i \ne j$. For another voter $k$, since $d(j,i)=d_k(j,i) \le d_k(j, y_k^*) + d_k(i,y_k^*)$, summing up over all $k \ne i,j$, we have
	$$(n-2)d(j,i)  \le \sum_{k\ne i,j}(d_k(j, y_k^*) + d_k(i,y_k^*)).$$
	
	Summing up over all $i,j$ such that $i<j$, we have
	\begin{align*}
		(n-2)\sum_{i<j}d(i,j)
		&\le \sum_{i<j}\sum_{k\ne i,j}
		\left(d_k(j,y_k^*) + d_k(i,y_k^*)\right) \\
		&=\sum_{k \in N}\sum_{i<j, \, i,j \ne k}
		\left(d_k(j,y_k^*) + d_k(i,y_k^*)\right) \\
		&=(n-2)\sum_{k\in N}\sum_{\ell\neq k} d_k(\ell,y_k^*),
	\end{align*}
	where the last identity holds since for each fixed $\ell \ne k$, $d_k(\ell, y_k^*)$ appears exactly once for every other voter in $N\setminus \{k, \ell\}$. Since there are $n-2$ such voters, we have the identity. 
	Dividing both sides by $n-2$ and re-indexing $k,\ell$ by $i,j$, we have
	$$\sum_{i<j}d(j,i) \le \sum_{i \ne j}d_i(j,y_i^*).$$
	
	Hence,
	$$ \frac{\sum_{i \in N}\sc_{d_{i}}(x_{i})}{\sum_{i \in N} \sc_{d_{i}}(y_{i}^*)} \le \frac{nS_0+2S_1}{S_0 + S_1},$$
	where $S_0 = \sum_{i \in N}d_i(i,y_i^*)$ and $S_1 = \sum_{i \ne j}d_i(j,y_i^*)$.
	
	Note that $d_i(i,y_i^*) \le d(i,j)+d_i(j,y_i^*)$. Summing over all $i,j$ such that $i \ne j$ yields
	$$(n-1)S_0 \le \sum_{i \ne j}d(i,j) + S_1 \le 3S_1.$$
	Hence, $S_0 \le \frac{3}{n-1}S_1$ holds, and the right-hand side above under this constraint is maximized when $S_0 = \frac{3}{n-1}S_1$:
	\begin{equation*}
		\frac{\sum_{i \in N}\sc_{d_{i}}(x_{i})}{\sum_{i \in N} \sc_{d_{i}}(y_{i}^*)} \le \frac{5n-2}{n+2} < 5. \qedhere
	\end{equation*}
\end{proof}

Now we complement the upper bounds with striking lower bounds. It turns out that in both adversarial models, persistent randomized dictatorship provides optimal distortions. We prove lower bounds of $3$ and $5$ for oblivious adversaries and adaptive adversaries, respectively.

We first consider adaptive adversaries.
The lower bound is constructed by two types of metric instances that the algorithm cannot distinguish from the observed information. One instance has the first candidate $c_1$ as the optimal candidate, and accepting challengers will accumulate the social costs. The other instance has one optimal challenger and not accepting it would lead to a high social cost. These two instances are in a trade-off: accepting many challengers is costly as it accumulates in the first instance, but accepting few will increase the chance of missing the optimal candidate in the latter instance.

\begin{theorem}
	Under adaptive adversaries, no randomized voting rule can guarantee a constant metric distortion below $5$ in the partial information IC model. 
\end{theorem}
\begin{proof}
	Fix $n \ge 3$ voters and horizon $T$.
	We first specify the potential candidates that can arrive. For every subset $S \subseteq \{2, \ldots, T\}$, create a potential candidate $v_S$. For a potential candidate $v_S$, let $h(v_S) = |S|$ be its depth.
	
	Set $c_1 = v_\emptyset$.
	At round $t \ge 2$, if the incumbent is $x_{t-1} = v_S$, introduce $c_t = v_{S \cup \{t\}}$.
	
	For an integer $h$, let $i_h = 1 + (h \bmod n)$. 
	We will construct two different metrics which both yield the following identical input to the algorithm: at any round, an incumbent of depth $h$ is preferred over the challenger only by voter $i_h$, and every other voter $i \ne i_h$ prefers the challenger.

	\paragraph{First metric.}	
	Let $u^0 = (1, \ldots, 1) \in \mathbb{R}^n$ and recursively define $u^{h+1} = u^h + 2 e_{i_h}$ where $e_i$ is the vector with $1$ only in coordinate $i$ and $0$ elsewhere. For a potential candidate $v$ with depth $h(v)=h$, define the voter-candidate distances by $d(i,v) = u^h_i$.
	Note that if the incumbent $x_{t-1}$ at round $t$ has depth $h$, then the challenger $c_t$ has depth $h+1$. Hence, we have 
	$$d(i_h, c_t) = u_{i_h}^{h+1} = u_{i_h}^h + 2 = d(i_h, x_{t-1}) + 2 > d(i_h, x_{t-1}),$$
	and voter $i_h$ prefers the incumbent over the challenger.
	For other voters, the distances to $x_{t-1}$ and $c_t$ are equal, and we break ties in favor of the challenger.
	Hence, this metric induces the described input.
	
	Note that this extends to a full metric. Fix a full metric, e.g., set all voter–voter distances to $2$.
	Again, it is sufficient to check for $3$-path triangle inequalities: $d(i,c) \le d(i,c') + d(i',c') + d(i',c)$ for all $i, i' \in N$ and $c, c' \in C$.
	For every $t$ and every pair of voters $i, i'$, by definition we have $d(i,c_t)\ge 1$ and $|d(i,c_t) - d(i', c_t)|\le 2$. Hence, $d(i,c) - d(i',c) \le 2 \le d(i,c') + d(i', c')$, and the $3$-path triangle inequality holds.

	Under this metric, the social cost of $v_S$ is $n + 2|S|$, and hence the initial candidate $c_1$ is always optimal with cost $n$. Hence, the metric distortion at round $T$ becomes 
	$$\frac{\Exp[\sc(x_T)]}{n} = 1 + \frac{2}{n}\Exp[h(x_T)].$$
	
	\paragraph{Second metric.}
	We then construct a different metric that induces the same preference profile.
	Fix a number $k \in \{2, \ldots, T\}$. Call a potential candidate $v_S$ good if $k \in S$ or bad otherwise. Hence, before round $k$, all candidates are bad, at round $k$, the challenger is good, and after round $k$, either both the incumbent and the challenger are good or both bad.
	
	For a voter $i \in N$, define $n$-vectors $b^i = (5, \ldots, 5) - 2 e_i$ and $g^i = (1, \ldots, 1) + 2 e_i$.
	Then define the voter-candidate distances for $v_S$ with $h(v_S) = h$ by
	$$
	d( i ,v_S)=
	\begin{cases}
		b^{i_h}_i, & k \notin S,\\
		g^{i_{h-1}}_i, & k \in S.
	\end{cases}
	$$
	
	To check it extends to a full metric, again observe that by definition we have $d(i,c_t)\ge 1$ and $|d(i,c_t) - d(i', c_t)|\le 2$ for every $t$ and every pair of voters $i, i'$. Hence, $d(i,c) - d(i',c) \le 2 \le d(i,c') + d(i', c')$, and the $3$-path triangle inequality holds.
	
	Break ties by preferring bad candidates to good candidates, and then preferring greater depth.
	We claim that this metric also induces the described input.
	For an incumbent of depth $h$, the challenger has depth $h+1$.
	If the incumbent is bad, it has distance-vector $b^{i_h}$ to the voters. Then, the challenger has distance-vector $g^{i_h}$ if it is round $k$ or $b^{i_{h+1}}$ otherwise.
	If the incumbent is good, it has distance-vector $g^{i_{h-1}}$ and the challenger has distance-vector $g^{i_h}$.
	In these three cases, voter $i_h$ has distances $3,3,1$ to the incumbent and $5,3,3$ to the challenger, respectively. Hence, $i_h$ always prefers the incumbent.
	Voter $i_{h-1}$ has distances $5,5,3$ to the incumbent and $5,1,1$ to the challenger, respectively. Hence, $i_{h-1}$ always prefers the challenger.
	Voter $i_{h+1}$ has distances $5,5,1$ to the incumbent and $3,1,1$ to the challenger, respectively. Hence, $i_{h+1}$ always prefers the challenger.
	Every other voter $i \notin \{i_{h-1}, i_h, i_{h+1}\}$ has distances $5,5,1$ to the incumbent and $5,1,1$ to the challenger, respectively. Hence, $i$ always prefers the challenger, and the described input is induced.
	
	Under this metric, the social cost of a bad candidate is $5n-2$ and that of a good candidate is $n+2$.
	A good candidate always arrives, and hence is always optimal with social cost $n+2$.
	Let us compute the expected social cost of the final round winner $x_T$.
	
	Since the algorithm cannot observe the hidden $k$, $x_T$ is independent of the choice of $k$.
	Let $p_k = \Pr(k \in S_T)$ for each $k \in \{2, \ldots, T\}$ where $x_T = v_{S_T}$. 
	Since $x_T$ is good if and only if $k \in S_T$, the expected social cost of $x_T$ becomes $p_k (n+2) + (1-p_k) (5n-2)$.
	Hence, the metric distortion at round $T$ becomes 
	$$\frac{\Exp[\sc(x_T)]}{n+2} = \frac{5n-2}{n+2} -  \frac{4n-4}{n+2} p_k.$$
	
	Now we show that we can choose $k$ to have small $p_k$.
	Since $x_T$ is good if and only if $k \in S_T$, we have 
	$$\sum_{k=2}^T \mathbf{1}(k \in S_T) = |S_T| = h(x_T).$$
	
	Taking expectations, we have $\sum_{k=2}^T p_k = \Exp[h(x_T)]$.
	Hence, there exists $k$ with $p_k \le \Exp[h(x_T)]/(T-1)$, and with this $k$, the metric distortion of $x_T$ is at least
	$$\frac{5n-2}{n+2} -  \frac{4n-4}{(T-1)(n+2)} \Exp[h(x_T)] \ge \frac{5n-2}{n+2} - \frac{4}{T-1}\Exp[h(x_T)].$$
	Note that this choice can be inferred solely from the algorithm's probability law, not its realized random choices.
	
	\paragraph{Choosing the metric.}
	Now take $T = n^2 + 1$. Using the algorithm's law, compute $\Exp[h(x_T)]$ with respect to the described input. If $\Exp[h(x_T)] \ge 2n$, choose the first metric to simulate the input, which gives distortion at least $1 + \frac{2}{n} \cdot 2n = 5$.
	If $\Exp[h(x_T)] < 2n$, choose the second metric, which gives distortion at least
	$$\frac{5n-2}{n+2} - \frac{4}{n^2} 2n \ge 5 - \frac{20}{n}.$$
	For any constant below $5$, the lower bound exceeds it for sufficiently large $n$.
\end{proof}

\begin{remark}
	The above lower bound instance still applies for the larger, full candidate space $C_t$.
	After any choice of incumbent $x_{t-1}=v_S$, including a return to an earlier candidate, the adversary introduces $c_t=v_{S\cup\{t\}}$. Thus every challenger remains a child of the actual incumbent, and all constructed metrics continue to produce identical incumbent–challenger comparisons. Since the distortion bounds are derived from the depth of the final winner, it remains valid regardless of the algorithm’s intermediate choices.
\end{remark}


Finally, we prove persistent randomized dictatorship is also optimal under oblivious adversaries. We again consider two instances that the algorithm cannot distinguish.
\begin{theorem}\label{thm:pi-ic-obliv}
	Under oblivious adversaries, no randomized voting rule can guarantee a constant metric distortion below $3$ in the partial information IC model. 
\end{theorem}
\begin{proof}
	Fix $n \ge 2$ and $D < 3 - \frac{2}{n}$. We show that no randomized voting rule can guarantee distortion at most $D$. Fix a randomized voting rule, a round $t \ge 2$, and a complete preference profile on the first $t-1$ candidates $c_1, \ldots, c_{t-1}$. Assume the algorithm guarantees distortion $D$.
	For each voter $i \in N$, let $X_i$ be the indicator random variable which indicates whether the algorithm retains the incumbent $x_{t-1}$ on round $t$ when exactly voter $i$ prefers $x_{t-1}$ over $c_t$.
	The randomness of $X_i$ comes from the algorithm (and the observations of actual previous winners).
	We first bound the expectation of $X_i$.
	
	Consider the instance on a line which places the first $t-1$ candidates at $0$, voter $i$ at $1$, and the challenger $c_t$ and all other voters at $2$. Break ties consistently so that voter $i$ prefers $c_s$ over $c_t$ for every $s < t$.
	The first $t-1$ candidates have social cost of $2n-1$ and $c_t$ has social cost of $1$.
	Then, at round $t$, the algorithm has distortion of
	$$\Pr(X_i = 0) \cdot 1 + \Pr(X_i=1) \cdot (2n-1) = 1 + 2(n-1) \Exp[X_i] \le D,$$
	or equivalently, $\Exp[X_i] \le \frac{D-1}{2(n-1)}$.
	
	Now we construct another instance.
	For each candidate $c$, associate it with a set $B(c) \subseteq [n]$.
	Define voter-candidate distances by
	$$d(i,c)=
	\begin{cases}
		1,&i\in B(c),\\
		3,&i\notin B(c).
	\end{cases}
	$$
	Setting voter-voter distances and candidate-candidate distances to $2$ yields a valid metric.
	Break ties in favor of late arrivals. Hence, at round $t$, the incumbent's supporters are exactly $B(x_{t-1}) \setminus B(c_t)$.

	Note that candidate $c$ has social cost $3n - 2|B(c)|$.
	Set $B(c_1) = [n]$ so that $c_1$ always has optimal social cost.
	Then since the algorithm has distortion guarantee $D$, for every round $t$, we have
	${\Exp[3n-2|B(x_t)|]}/{n} \le D$, or equivalently,
	$$\Exp[|B(x_t)|] \ge \frac{n(3-D)}{2} > 1.$$
	We claim we can select the sets $B(c_t)$ for $t \ge 2$ in a specific manner leading to a contradiction.
	
	We show such existence using the probabilistic method.
	For a parameter $\lambda > 1$, independently sample each $B(c)$ for $c \ne c_1$ according to
	$$\Pr(B(c) = S) = \frac{\lambda^{-|S|}}{Z},$$
	where $S \subseteq [n]$ and $Z=(1+\lambda^{-1})^n$. These sets correspond to the randomness of the instance.
	
	For a nonnegative integer $k$, let $\phi(k) = \frac{Z}{\lambda - 1}\lambda^k$. (Let $\phi(-1)=0$.)
	Let $b_t = |B(x_t)|$. We will keep track of the expected change in the potential $\phi(b_t)$. 
	At round $t\ge 2$, if $x_t = x_{t-1}$, we have $\phi(b_t) = \phi(b_{t-1})$. 
	If $x_t = c_t$, the increment $\phi(b_t) - \phi(b_{t-1})$ is bounded by $\phi(b_t) = \phi(|B(c_t)|)$.
	Since $c_t \ne c_1$, we have $\Exp[\phi(|B(c_t)|)] = 2^n / (\lambda - 1) =: \eta > 0$.
	
	There are also cases where $\phi(b_t)$ decreases. 
	Fix a round $t$.
	Let $B_{<t} = (B(c_s))_{2 \le s \le t-1}$ represent the randomness of the instance until round $t-1$. We will condition on $B_{<t}$.
	
	For each $i \in N$, consider the event $E_i = \{i \in B(x_{t-1}),\ B(c_t) = B(x_{t-1}) \setminus \{i\}\}$, and let $E = \cup_{i \in N} E_i$. The events $E_i$ are disjoint as for $i,j \in N$, $i,j \in B(x_{t-1})$ and $B(x_{t-1}) \setminus \{i\} = B(x_{t-1}) \setminus \{j\}$ implies $i = j$. On event $E_i$, since $B(x_{t-1}) \setminus B(c_t) = \{i\}$, that is, exactly voter $i$ prefers the incumbent, the algorithm selects the incumbent if and only if $X_i = 1$.
	Note that $B(x_{t-1})$ and $b_{t-1}$ may still be random conditional on $B_{<t}$.
	Since $B(c_t)$ is independent of $B_{<t}$ and the algorithm's randomness, we have
	\begin{align*}
		&\Exp[(\phi(b_t) - \phi(b_{t-1})) \mathbf{1}_{E_i} | B_{<t}] \\
		&=  \Exp[(\phi(b_{t-1}-1) - \phi(b_{t-1})) \mathbf{1}(\{i \in B(x_{t-1})\}) \mathbf{1}(B(c_t) = B(x_{t-1}) \setminus \{i\}) (1 - X_i)| B_{<t}] \\
		&= \Exp\left[ (\phi(b_{t-1}-1) - \phi(b_{t-1}))\mathbf{1}(\{i \in B(x_{t-1})\}) \frac{\lambda^{-(b_{t-1}-1)}}{Z} (1-X_i)  \middle| B_{<t} \right] \\
		&= -\Exp\left[ \mathbf{1}(\{i \in B(x_{t-1})\})(1-X_i) \middle| B_{<t} \right],
	\end{align*}
	where the last identity follows from $\Pr(B(c) = S) (\phi(|S|+1) - \phi(|S|)) = 1$.
	
	Hence, 
	\begin{align*}
		\Exp[(\phi(b_t) - \phi(b_{t-1})) \mathbf{1}_{E} | B_{<t}] &= \sum_{i \in N} \Exp[(\phi(b_t) - \phi(b_{t-1})) \mathbf{1}_{E_i}| B_{<t}]  \\
		&= - \Exp[b_{t-1} | B_{<t}] +\Exp\left[  \sum_{i \in B(x_{t-1})} X_i \middle| B_{<t} \right].
	\end{align*}
	
	Also, from previous analysis, $\Exp[(\phi(b_t) - \phi(b_{t-1})) \mathbf{1}_{E^c} | B_{<t}] \le \Exp[\phi(|B(c_t)|)] = \eta$. Together, these yield 
	$$\Exp[\phi(b_t) - \phi(b_{t-1}) | B_{<t}] \le - \Exp[ b_{t-1} | B_{<t}] + \Exp \left[\sum_{i \in B(x_{t-1})} X_i \middle| B_{<t} \right] + \eta.$$
	Since $X_i \ge 0$, we have	
	$$\Exp\left[\sum_{i \in B(x_{t-1})} X_i \middle| B_{<t} \right]  \le \sum_{i=1}^n \Exp[X_i | B_{<t}] \le \frac{n(D-1)}{2(n-1)} =: \beta < 1,$$
	where the second inequality holds from the argument above.
	Now taking the expectation over $B_{<t}$, by the tower property, we have
	$$\Exp[\phi(b_t) - \phi(b_{t-1})] \le - \Exp[b_{t-1}] + \beta + \eta.$$
	Take sufficiently large $\lambda$ so that $\delta = 1- \beta - \eta > 0$. Since $\Exp[b_{t-1}] > 1$, we have $\Exp[\phi(b_t) - \phi(b_{t-1})] \le -\delta < 0$. Since $b_1 = n$, summing over rounds, we have $\Exp[\phi(b_T)] \le \phi(n) - (T-1) \delta$. Taking sufficiently large $T$ yields $\Exp[\phi(b_T)] < 0$, contradicting that $\phi$ is nonnegative.
\end{proof}

\begin{remark}
	We can again extend the result to full candidate space by generalizing the proof. We leave the details in \Cref{sec:ic-fi}.
\end{remark}

\section{Incumbent-Challenger Model with Full Information}\label{sec:ic-full-info}

Now we consider the IC model under full information. At round $t$, we observe the full ranking of voters on $C_t = \{c_1, \ldots, c_t\}$. 

\subsection{Deterministic Voting Rules}
Can deterministic voting rules achieve constant metric distortion under this model?
One can observe that Plurality Veto~\citep{kizilkaya2022plurality} cannot be applied directly even with full information, as it can select previously discarded candidates.
\begin{example}
	Consider the case with $T=3$ and $3$ candidates $a,b,c$ arriving in this order. There are nine voters, with the offline ranking being
	\[
	\begin{array}{c|c}
		\text{\# of voters} &\text{Ranking} \\
		\hline
		2 &
		a \succ b \succ c
		\\[2mm]
		4
		& b \succ c \succ a
		\\[2mm]
		3
		& c \succ a \succ b
	\end{array}
	\]
	Run Plurality Veto each round. On round $t=2$, $a$ wins as 5 voters have $a \succ b$, and $b$ is discarded. However, on round $t=3$, plurality scores are $s(a)=2$, $s(b) =4$, $s(c) = 3$, and regardless of the order of the voters, $a$ is vetoed first. Then since there are six voters who rank $b$ over $c$, $c$ is vetoed next. Hence, the Plurality Veto winner is $b$, which cannot be selected under the incumbent-challenger model.
\end{example}

Still, we show that constant distortion is possible. 
Copeland, which has a metric distortion of 5~\citep{anshelevich2015metric}, is again incompatible with the incumbent-challenger model as it might choose previously discarded candidates.
However, the essential analysis that yielded the metric distortion bound was from the fact that Copeland selects a candidate that is \emph{uncovered} in the tournament graph. \citet{munagala2019weighted} extended this observation to consider \emph{weighted tournament graphs}, improving the metric distortion bound to $2+\sqrt{5}$.
We show that even this extended version can be made compatible with the IC model, achieving metric distortion of at most $2+\sqrt{5}$ for each round.

We first introduce some notation.
Fix $\lambda \in [\frac{1}{2}, 1]$. Given an election instance, a candidate $c \in C$ is in the \emph{$\lambda$-weighted uncovered set}~\citep{munagala2019weighted} if for each candidate $c' \ne c$, 
\begin{itemize}
	\item either $|\{i \in N: c \succ_i c'\}| \ge (1-\lambda) |N|$,
	\item or there exists $c'' \notin \{c, c'\}$ such that $|\{i \in N: c \succ_i c''\}| \ge (1-\lambda) |N|$ and $|\{i \in N: c'' \succ_i c'\}| \ge \lambda |N|$.
\end{itemize}

\begin{lemma}[\citealp{munagala2019weighted}]\label{lem:munagala}
	Let $\varphi = \frac{\sqrt{5} -1}{2} \approx 0.618$. A candidate in the $\varphi$-weighted uncovered set always exists and has metric distortion at most $2 + \sqrt{5}$.
\end{lemma}

\begin{algorithm}[t]
	\begin{algorithmic}
		\State \textbf{Input}: An online voting instance of $T$ rounds
		\State \textbf{Output}: A sequence of winners $x_1,\ldots,x_T$ satisfying $x_t\in\{x_{t-1},c_t\}$ for every $t\ge 2$
		\State $\varphi \gets \frac{\sqrt{5}-1}{2}$
		\State $x_1 \gets c_1$
		\For{$t=2,\ldots,T$}
		\If{$x_{t-1}$ is in the $\varphi$-weighted uncovered set of $E_t=(N,C_t,\succ|_{C_t \times C_t})$}
		\State $x_t \gets x_{t-1}$
		\Else
		\State $x_t \gets c_t$
		\EndIf
		\EndFor
		\State \Return $(x_1,\ldots,x_T)$
	\end{algorithmic}
	\caption{Online weighted-uncovered-set rule}
	\label{alg:online-weighted-uncovered}
\end{algorithm}

This allows us to design a deterministic rule with constant metric distortion.

\begin{theorem}
	\Cref{alg:online-weighted-uncovered} achieves metric distortion at most $2+\sqrt{5}$ in the full information IC model.
\end{theorem}
\begin{proof}
	By \Cref{lem:munagala}, it is sufficient to show that the winner $x_t$ at round $t$ is in the $\varphi$-weighted uncovered set of $E_t=(N,C_t,\succ|_{C_t \times C_t})$. 
	We prove this by induction on $t$. The case is trivial for $t=1$ or $t=2$ as the $\varphi$-weighted uncovered set is nonempty. 
	Assume $x_{t-1}$ is in the $\varphi$-weighted uncovered set of $E_{t-1}$. 
	If $x_{t}=x_{t-1}$, by definition, $x_{t}$ is in the $\varphi$-weighted uncovered set of $E_t$.
	On the other hand, if $x_t = c_t$, then $x_{t-1}$ is not in the $\varphi$-weighted uncovered set of $E_t$. Since the addition of $c_t$ does not change the pairwise rankings between candidates in $C_{t-1}$, it must be that $c_t$ blocked $x_{t-1}$ from being in the $\varphi$-weighted uncovered set, i.e.,
	\begin{enumerate}[label=(\arabic*)]
		\item\label{item:1} $|\{i \in N: x_{t-1} \succ_i c_t\}| < (1-\varphi) |N|$ and
		\item\label{item:2} for each $c' \in C_t \setminus \{x_{t-1}, c_t\}$, $|\{i \in N: x_{t-1} \succ_i c'\}| < (1-\varphi) |N|$ or $|\{i \in N: c' \succ_i c_t\}| < \varphi |N|$.
	\end{enumerate}
	
	Now let us show that $c_t$ is in the $\varphi$-weighted uncovered set of $E_t$. For $x_{t-1} \in C_t$, since $|\{i \in N: c_t \succ_i x_{t-1}\}| > \varphi |N| \ge (1-\varphi)|N|$, it does not block $c_t$.
	For $c' \in C_t \setminus \{x_{t-1}, c_t\}$, since $x_{t-1}$ was in the $\varphi$-weighted uncovered set of $E_{t-1}$,
	\begin{itemize}
		\item\label{case:1} either $|\{i \in N: x_{t-1} \succ_i c'\}| \ge (1-\varphi) |N|$,
		\item\label{case:2} or there exists $c'' \in C_{t-1} \setminus \{x_{t-1}, c'\}$ such that $|\{i \in N: x_{t-1} \succ_i c''\}| \ge (1-\varphi) |N|$ and $|\{i \in N: c'' \succ_i c'\}| \ge \varphi |N|$.
	\end{itemize}
	For the former case, by \labelcref{item:2}, we have $|\{i \in N: c_t \succ_i c'\}| \ge (1-\varphi)|N|$. For the latter case, since $|\{i \in N: x_{t-1} \succ_i c''\}| \ge (1-\varphi) |N|$ for some $c''$, again by \labelcref{item:2}, we have $|\{i \in N: c_t \succ_i c''\}| \ge (1-\varphi)|N|$. Since $|\{i \in N: c'' \succ_i c'\}| \ge \varphi |N|$ in the latter case, $c'$ does not block $c_t$.
	Hence, $x_t$ is in the $\varphi$-weighted uncovered set of $E_t$.
\end{proof}

Hence, a constant distortion can be attained with a deterministic voting rule throughout the entire process, unlike the partial information case. In fact, the constant $2 + \sqrt{5}$ is optimal.

\begin{theorem}\label{thm:FI-IC-det-lb}
	No deterministic voting rule can guarantee a constant metric distortion below $2 + \sqrt{5}$ in the full information IC model.
\end{theorem}

The proof again considers for each $\theta \in (0,1)$, one ordinal preference profile which has multiple metrics consistent with it. A deterministic voting rule either always accepts the challenger each round, or rejects some challenger at some round.
One metric will yield the former case distortion of $\frac{1 + \theta}{1 - \theta}$, and another metric will yield distortion of $1 + \frac{2}{\theta}$. Thus, maximizing $\min \left\{ 1 + \frac{2}{\theta} , \frac{1 + \theta}{1 - \theta}\right\}$ over $\theta \in (0,1)$, we obtain distortion $2 + \sqrt{5}$ by $\theta = (\sqrt{5}-1)/2$.

We first prove a lemma that allows us to construct consistent metrics from certain preference profiles.
\begin{lemma}\label{lem:metric-realization}
	Let a preference profile $\succ$, nonnegative rational heights $(h_c)_{c\in C}$ of the candidates, and nonnegative rational numbers $(\delta_i)_{i \in N}$ be given. 
	For voter $i \in N$, let $F_i(c)=\min \{h_{c'}: c' \precsim_i c \}$ be the minimum height among candidates preferred at most $c$.
	If
	\begin{enumerate}[label=(\arabic*)]
		\item $\sum_{i \in N}\delta_i >0$ and for each voter $i \in N$, $c \succ_i c'$ implies $h_c-h_{c'}\le\delta_i$, and
		\item there exists a candidate $y$ with $h_y = 0$,
	\end{enumerate}
	then there exists a consistent pseudometric $d$ such that
	$$\frac{\sc(x)}{\sc(y)}=1+2\frac{\sum_{i \in N} F_i(x)}{\sum_{i \in N}\delta_i},$$
	for each candidate $x \in C$.
\end{lemma}
\begin{proof}
	We construct the desired pseudometric $d$.
	Define voter-candidate distances by $d(i,c)={\delta_i}/{2}+F_i(c)$.
	For distinct candidates $c, c'$, let $d(c,c') = \max\{h_c,h_{c'}\}$.
	For distinct voters $i,j$, define $d(i,j)$ by the shortest path distance: $d(i,j)=\min_{c\in C}\{d(i,c)+d(j,c)\}$.
	To show this is a valid metric, observe that for distinct candidates $a,b,c \in C$, $d(a,c) = \max\{h_a, h_c\} \le \max\{h_a, h_b, h_c\} \le d(a,b) + d(b,c)$.
	It hence suffices to show $|d(i,c) - d(i,c')| \le d(c,c') \le d(i,c) + d(i,c')$ for every $i \in N$ and $c,c' \in C$.
	This is because voter-voter distances are defined by shortest path distances.
	
	Observe that, by the assumptions, $\max\{0,h_c-\delta_i\}\le F_i(c)\le h_c$.
	Hence, for any voter $i$ and distinct candidates $c,c'$,
	$$|d(i,c)-d(i,c')| = |F_i(c)-F_i(c')| \le d(c,c').$$
	Moreover, assuming $h_c \ge h_{c'}$, we have
	$$d(i,c)+d(i,c') =  \delta_i + F_i(c) + F_i(c') \ge \delta_i + F_i(c) \ge h_c = d(c,c').$$
	Hence, the pseudometric is valid.
	
	Since $c \succ_i c'$ implies $F_i(c) \le F_i(c')$ by definition, the resulting pseudometric is consistent with $\succ$. Finally, since $h_y=0$, we have $F_i(y)=0$ for every voter $i$. Therefore, for every candidate $x$,
	\begin{equation*}
		\frac{\sc(x)}{\sc(y)} = \frac{\sum_{i\in N}(\delta_i/2+F_i(x))}{\sum_{i\in N}\delta_i/2} 
		= 1 + 2 \frac{\sum_{i\in N}F_i(x)}{\sum_{i\in N}\delta_i}. \qedhere
	\end{equation*}
\end{proof}

Now we proceed to prove the lower bound.

\begin{proof}[Proof of \Cref{thm:FI-IC-det-lb}]
	Fix a deterministic voting rule, a rational number $\theta \in (0,1)$, and an integer $L \ge 2$. Let $T=L(L-1)+1$, and consider $T$ candidates arriving in the order $c_1,c_2,\ldots,c_T$.
	
	We construct two types of voters. A fraction $1-\theta$ of the voters, denoted by type $A$, prefers candidates arriving later: $c_T \succ c_{T-1} \succ \cdots \succ c_1$.
	The remaining fraction $\theta$ is divided equally into $L$ groups $B_1, \ldots, B_L$. For each $\ell \in [L]$, voters in $B_\ell$ partition the candidates into blocks according to
	$$b_\ell(t)=\left\lfloor\frac{t+\ell-2}{L}\right\rfloor.$$
	That is, candidate $c_t$ is in block $b_\ell (t)$.
	The voters prefer candidates in earlier blocks to those in later blocks, but within the same block, they prefer candidates arriving later. Note that since $\theta$ is rational, all voter fractions can be realized by a finite voter set $N$. Fix one realization and let $n$ be the number of voters.
	
	Observe that for every pair of consecutive candidates $c_{t-1},c_t$, exactly one group $B_\ell$ prefers $c_{t-1}$ to $c_t$, while all other voters (including type $A$) prefer $c_t$. Thus,
	\begin{equation}\label{eq:chain-support}
		\frac1n|\{i\in N:c_t\succ_i c_{t-1}\}| =1-\frac{\theta}{L}.
	\end{equation}
	On the other hand, all voters in the groups $B_\ell$ prefer $c_1$ to $c_T$.
	
	We consider two cases depending on how the voting rule behaves on this preference profile.
	
	\paragraph{Case 1: The rule rejects some challenger.}
	Let $t$ be the first round the rule selects the incumbent. Note that $x_{t-1} = c_{t-1}$. Consider the preference profile restricted to $C_t$. Assign heights $h_{c_t}=0$ and $h_{c_s}=1$ for $s<t$, and set $\delta_i=0$ for $i\in A$ and $\delta_i=1$ otherwise.
	
	These values satisfy the assumptions of \Cref{lem:metric-realization}. We have $\sum_{i \in N} \delta_i=n\theta > 0$. Moreover, since a voter $i \in A$ ranks $c_t$ first among the available candidates, $c \succ_i c'$ implies $h_c \le h_{c'}$. The inequality trivially holds for voters of other types.
	
	Also, since $F_i(c_{t-1})=1$ if and only if $c_t \succ_i c_{t-1}$, by \eqref{eq:chain-support}, we have $\sum_{i \in N} F_i(c_{t-1}) = n\left(1-{\theta}/{L}\right)$.
	Applying \Cref{lem:metric-realization} with $y=c_t$, we obtain distortion at round $t$ at least
	\begin{equation}\label{eq:reject-bound}
		\frac{\sc(c_{t-1})}{\sc(c_t)} =1+2\frac{n(1-\theta/L)}{n\theta} = 1+ \frac{2}{\theta} - \frac{2}{L}.
	\end{equation}
	For the remaining rounds, place challengers arbitrarily as we already obtained necessary bounds. The preferences of remaining candidates need not match the above instance.
	
	\paragraph{Case 2: The rule accepts every challenger.}
	In this case, the final-round winner is $x_T = c_T$. Assign heights $h_{c_t}=(t-1)/(T-1)$, and set $\delta_i=1$ for $i\in A$ and $\delta_i=1/L$ otherwise.
	
	These values again satisfy \Cref{lem:metric-realization}. Indeed, for voters in $A$, all height differences are at most $1$. For voters in $B_\ell$, heights decrease along their rankings only within blocks, where the decrease is at most $(L-1)/(T-1)=1/L$. Also, $\sum_{i \in N} \delta_i =n\left(1-\theta+\frac{\theta}{L}\right) > 0$.

	Now we compute the distortion at round $T$.
	For a voter $i$ of type $A$, we have $F_i(c_T)= \min_{t} h_{c_t} = 0$. For a voter $i$ of type $B_\ell$, the last block starts at $c_{T-\ell+1}$, so $F_i(c_T)=(T-\ell)/(T-1)$. Consequently,
	$$\sum_i F_i(c_T) = \frac{n\theta}{L} \sum_{\ell=1}^{L} \frac{T-\ell}{T-1} =n\theta\left(1-\frac1{2L}\right).$$
	Applying \Cref{lem:metric-realization} with $y=c_1$, we obtain
	\begin{equation}\label{eq:accept-bound}
		\begin{aligned}
			\frac{\sc(c_T)}{\sc(c_1)}
			&=1+2\frac{\theta(1-1/(2L))}
			{1-\theta+\theta/L}\\
			&=\frac{1+\theta}{1-\theta+\theta/L}.
		\end{aligned}
	\end{equation}
	
	\paragraph{Combining the cases.}
	By \eqref{eq:reject-bound} and \eqref{eq:accept-bound}, the worst-case distortion of the rule is at least $\min\left\{ 1 + \frac{2}{\theta} - \frac{2}{L}, \frac{1+\theta}{1-\theta+\theta/L} \right\}$.
	Taking $L\to\infty$, this approaches $\min \left\{ 1 + \frac{2}{\theta} , \frac{1 + \theta}{1 - \theta}\right\}$.
	Thus, maximizing above over $\theta \in (0,1)$, we obtain $2 + \sqrt{5}$ by $\theta^* = (\sqrt{5}-1)/2$.

	Hence, for a constant $D$ below	$2+\sqrt{5}$, we can take sufficiently large $L$ and a rational number $\theta$ sufficiently close to $\theta^*$ so that the distortion is above $D$.
	Since the rule is deterministic, the adversary can determine in advance whether and when the rule first rejects a challenger. It can therefore fix the corresponding complete instance and metric before execution, actually yielding distortion over $D$.
\end{proof}

\begin{remark}
	In the partial information setting, we cannot implement \Cref{alg:online-weighted-uncovered} as we cannot determine whether $x_{t-1}$ is in the $\varphi$-weighted uncovered set without the information between $c_t$ and $C_{t-1}\setminus \{x_{t-1}\}$.
\end{remark}

\subsection{Randomized Voting Rules}
\Cref{alg:persistent-random-dictatorship} surely works for the full information IC model, and guarantees the same distortion bounds of $3$ and $5$ against oblivious and adaptive adversaries, respectively. 
Will persistent randomized dictatorship again be optimal?

Let us first consider oblivious adversaries. It turns out that \Cref{alg:persistent-random-dictatorship} is again optimal in this case.
\begin{theorem}
	Under oblivious adversaries, no randomized voting rule can guarantee a constant metric distortion below $3$ in the full information IC model. 
\end{theorem}
\begin{proof}
	Fix a randomized voting rule and $\epsilon > 0$. We construct an oblivious instance on which its expected distortion at the final round is at least $3 - \epsilon$.
	
	\paragraph{Instance.}
	Consider a perfect binary tree of depth $d$ with root $v_0$ whose leaves represent candidates, arriving from left to right. 
	Hence, there are a total of $T = 2^d$ rounds.
	We prove existence of our desired instance using probabilistic methods.
	For each internal node $v$, we consider a child pointer $p_v \in \{L,R\}$ that either points to the left child or the right child. 
	Given a pointer profile $p = (p_v)_v$, let $c_p(v)$ denote the leaf obtained by following these pointers from $v$.
	For each $p = (p_v)_v$ we will construct an instance in which $c^*_p = c_p(v_0)$ is the optimal candidate at the final round. We sample each pointer $p_v$ independently and uniformly from $\{L,R\}$ and show that some realization yields the desired instance.
	
	We now construct the instances for each $p=(p_v)_v$. 
	Enumerate the internal nodes of the tree by $v_1,\ldots,v_m$, where $m=2^d-1$.
	Fix an integer $q\ge 3$ so that $\alpha=1/q < \epsilon/4$. These are common across all profiles $p$.
	Construct a set of $n=q^m$ voters, indexed by $i = (i_1, \ldots, i_m) \in [q]^m$. For each voter $i$ and internal node $v_j$, define $X_{i,v_j}=1$ if $i_j=1$, and $X_{i,v_j}=0$ otherwise.
	Thus, for a uniformly random voter $i$ sampled from $N$, the values $(X_{i,v})_v$ are i.i.d. Bernoulli random variables with success probability $\alpha$. 
	
	\begin{wrapstuff}[r,width=5.5cm,type=figure]
		\begin{adjustbox}{max width=\linewidth}
			\begin{tikzpicture}[
				font=\small,
				vertex/.style={circle, draw, minimum size=6mm, inner sep=0pt}
				]
				
				\node[vertex] (v) at (0,3.3) {$v$};
				\node[vertex] (l) at (-1.55,2.35) {$\ell$};
				\node[vertex] (r) at (1.55,2.35) {$r$};
				
				\draw (v) -- (l);
				\draw (v) -- (r);
				
				\begin{scope}[shift={(-1.55,0)}]
					\filldraw[fill=gray!11, draw=gray!70]
					(0,2.14) -- (-0.7,1.05) -- (0.7,1.05) -- cycle;
					
					\draw[red!70!black, thick, ->]
					(0,2.10) -- (-0.08,1.86) -- (0.12,1.65)
					-- (-0.17,1.43) -- (0.23,1.23) -- (0.23,1.06);
					\fill[red!70!black] (0.23,1.06) circle (1.4pt);
					\node[red!70!black, font=\scriptsize, anchor=north]
					at (0.23,1.05) {$c_p(\ell)$};
				\end{scope}
				
				\begin{scope}[shift={(1.55,0)}]
					\filldraw[fill=gray!11, draw=gray!70]
					(0,2.14) -- (-0.7,1.05) -- (0.7,1.05) -- cycle;
				\end{scope}
				
				\node[font=\scriptsize] at (-1.55,0.4) {Ranking in $C_\ell$};
				\node[font=\scriptsize] at (1.55,0.4) {Ranking in $C_r$};
				\node[font=\scriptsize] at (-1.55,0.1)
				{$L^+ \succ c_p(\ell) \succ L^-$};
				\node[font=\scriptsize] at (1.55,0.1) {$R$};
				
				\node at (0,-0.7) {
					$\begin{aligned}
						X_{i,v}=1 &\implies R \succ L^+ \succ c_p(\ell) \succ L^-\\
						X_{i,v}=0 &\implies L^+ \succ c_p(\ell) \succ R \succ L^-
					\end{aligned}$
				};
			\end{tikzpicture}%
		\end{adjustbox}
	\end{wrapstuff}
	We now construct the preference ranking of each voter $i$, recursively. 
	Fix a pointer profile $p$.
	Recall that the candidates are the leaves. For a node $v$, write $T_v$ for the subtree rooted at $v$ and $C_v$ for the leaves of $T_v$.
	For a leaf, the ranking consists only of that candidate.
	At an internal node $v$ with left and right children $\ell, r$, suppose the rankings within the two child subtrees' leaves have already been constructed. If $X_{i,v}=1$, rank all of $C_r$ above all of $C_\ell$, preserving their respective rankings. Otherwise, insert the entire ranking of $C_r$ immediately below $c_p(\ell)$ in the ranking of $C_\ell$, again preserving the rankings within both subtrees. The ranking at the root defines the complete preference ranking $\succ_i$ of voter $i$.	
	
	Observe that the rankings restricted to $C_v$ and all candidates arriving before $C_v$ do not depend on $p_v$. Consequently, $p_v$ remains independent of the algorithm's observations and decisions through the round in which the last candidate in $C_v$ arrives.
	
	\paragraph{Bounding preference ratio.}
	As noted before, we will later construct a metric so that $c^*_p = c_p(v_0)$ yields the optimal social cost. In order to bound the distortion of the algorithm, let $S(x,y) = \frac{1}{n}|\{i \in N: x \succsim_i y\}|$ be the fraction of voters ranking $x$ at least as high as $y$. Equivalently, $S(x,y) = \Pr_{i \sim \operatorname{Unif}(N)}(x \succsim_i y)$. We will show that $\Exp[S(x_T, c^*_p)]$ can be arbitrarily small, and together with the specific metric we consider later, this will yield a desired distortion.
	Note that for $S(x,y)$, the following holds from the construction of the instances.
	For any leaf $x$, $S(x,x) = 1$. For any internal vertex $v$ with left and right children $\ell, r$ and a leaf $x \in C_v$,
	\[
	S(x,c_p(v))=
	\begin{cases}
		S(x,c_p(\ell)), & \text{if $x \in C_\ell$ and $p_v = L$},\\
		S(x,c_p(r)), & \text{if $x \in C_r$ and $p_v = R$},\\
		(1 - \alpha) S(x,c_p(\ell)), & \text{if $x\in C_\ell$ and $p_v = R$},\\
		\alpha, & \text{if $x\in C_r$ and $p_v = L$}.
	\end{cases}
	\]
	The first two cases are trivial by definition. If $x \in C_\ell$ and $p_v= R$, then since $c_p(v)=c_p(r)$, for voter $i$ to rank $x$ above $c_p(r) \in C_r$, it is equivalent to $X_{i,v} = 0$ and $x$ being ranked at least as high as $c_p(\ell)$. Hence, while keeping a pointer realization $p$ fixed,
	\begin{align*}
		S(x,c_p(v)) 
		&= \Pr_{i \sim \operatorname{Unif}(N)}(X_{i,v}= 0, x \succsim_i c_p(\ell)) \\
		&=  \Pr_{i \sim \operatorname{Unif}(N)}(X_{i,v}= 0) \Pr_{i \sim \operatorname{Unif}(N)}(x \succsim_i c_p(\ell)) \\
		&= (1- \alpha) S(x, c_p(\ell)).
	\end{align*}
	Similarly, if $x \in C_r$ and $p_v= L$, voter $i$ ranks $x$ above $c_p(v)=c_p(\ell) \in C_\ell$ if and only if $X_{i,v} = 1$. Hence, $S(x,c_p(v)) = \Pr_{i \sim \operatorname{Unif}(N)}(X_{i,v}= 1) = \alpha$.
	
	Consequently, if the path from the root $v_0$ to a leaf $x$ has a case where it moves in the opposite direction of the pointer $p$, if the path moves from a parent to the left child while the pointer points to the right child, a factor of $(1- \alpha)$ accumulates. Call this case a \emph{soft error}. If the path moves to the right child while the pointer points to the left, then $S(x, c_p^*) \le \alpha$. Call this case a \emph{fatal error}.
	If no fatal error occurs and there are $r$ soft errors, $S(x, c_p^*) = (1-\alpha)^r$.
	
	\paragraph{Survival probability.}
	Consider each $p_v$ as an independent, uniformly random variable over $\{L,R\}$.
	For a vertex $v$, let $x_v$ be the winner after the algorithm processes the round of the final candidate in $T_v$. For a nonnegative integer $k$, let $E_{v,k}$ be the event that $x_v \in C_v$ and the path from $v$ to $x_v$ contains no fatal errors and at most $k$ soft errors. 
	Note that $E_{v,k}$'s randomness comes from the algorithm's randomness (running on the entire tree) and the randomness of $p$.
	
	For nonnegative integers $j,k$, let $M_{j,k}$ be the supremum of $\Pr(E_{v,k})$ over all randomized algorithms and vertices $v$ with $\operatorname{depth}(T_v)=j$. We claim that for fixed $k$, $M_{j,k} \to 0$ as $j \to \infty$.
	
	Set $M_{0,k}=1$ for $k\ge0$ and $M_{j,-1}=0$.
	We first show the following recurrence:
	$$M_{j,k}\le M_{j-1,k}-\frac12 M_{j-1,k}^2+\frac12 M_{j-1,k-1}.$$
	
	To see this, consider an arbitrary voting rule and a vertex $v$ with $\operatorname{depth}(T_v)=j$. Let $\ell$ and $r$ be the left and right children of $v$.
	
	We decompose the probability $\Pr(E_{v,k})$:
	\begin{align*}
		\Pr(E_{v,k}) = \Pr&(E_{v,k} | E_{\ell, k-1})\Pr(E_{\ell, k-1})   \\
		& + \Pr(E_{v,k} | E_{\ell, k} \setminus E_{\ell, k-1}) \Pr(E_{\ell, k} \setminus E_{\ell, k-1}) + \Pr(E_{v,k} | E_{\ell, k}^c)\Pr(E_{\ell, k}^c).
	\end{align*}

	Note that $\Pr(E_{\ell,k})\le M_{j-1,k}$ and $\Pr(E_{\ell,k-1})\le M_{j-1,k-1}$.
	Let us bound each conditional probability.
	In $E_{\ell,k-1}$, we simply bound by $\Pr(E_{v,k} | E_{\ell, k-1}) \le 1$.
	
	In  $E_{\ell, k} \setminus E_{\ell, k-1}$, there are two possibilities for the possible winner $x_v$. First, if $x_v \in C_\ell$, then since all subsequent challengers belong to $C_r$, it must be that $x_v = x_\ell$. In this case, if $p_v = L$, no additional soft error occurs and $E_{v,k}$ holds. However if $p_v = R$, the path has $k+1$ soft errors and $E_{v,k}$ does not hold.
	Second, if $x_v \in C_r$, if $p_v = L$, a fatal error occurs and $E_{v,k}$ does not hold. Hence, in either case, the probability of $E_{v,k}$ occurring given $E_{\ell,k} \setminus E_{\ell, k-1}$ is at most $1/2$.
	
	In $E_{\ell, k}^c$, it must be that $x_v \ne x_\ell$ for $E_{v,k}$ to hold. Hence, $x_v \in C_r$ must hold, and then it must be $p_v = R$ to avoid fatal error. With $p_v = R$, $E_{v,k}$ happens if and only if $E_{r,k}$ holds. Hence, 
	$$\Pr(E_{v,k} | E_{\ell, k}^c) = \Pr(E_{r,k}, p_v = R | E_{\ell, k}^c) = \frac{1}{2}\Pr(E_{r,k}|E_{\ell,k}^c) \le \frac{1}{2}M_{j-1,k}.$$
	Here the second identity holds because $p_v$ is jointly independent of $E_{\ell,k}$ and $E_{r,k}$, while the last inequality follows because the pointers within $T_r$ are independent of the preceding history.
	
	Therefore,
	\begin{align*}
		\Pr(E_{v,k}) &\le \Pr(E_{\ell, k-1}) + \frac{1}{2} \left(\Pr(E_{\ell, k}) - \Pr(E_{\ell, k-1})\right) + \frac{1}{2}M_{j-1,k}(1-\Pr(E_{\ell, k}))\\
		&=\frac{1}{2}\Pr(E_{\ell,k-1}) +\frac{1}{2} M_{j-1,k} +\frac{1}{2} \Pr(E_{\ell,k})(1-M_{j-1,k}) \\
		&\le \frac{1}{2} M_{j-1,k-1} +\frac{1}{2} M_{j-1,k} + \frac{1}{2} M_{j-1,k}(1-M_{j-1,k}) \\
		&= M_{j-1,k} - \frac{1}{2} M_{j-1,k}^2 + \frac{1}{2} M_{j-1,k-1},
	\end{align*}
	where the second inequality follows from $\Pr(E_{\ell,k-1})\le M_{j-1,k-1}$ and $\Pr(E_{\ell,k})\le M_{j-1,k}$. 
	
	We now prove our claim.
	We perform induction on $k$, starting with $M_{j,-1}=0$. Assuming $M_{j,k-1} \to 0$ as $j \to \infty$, let $M = \limsup_{j\to\infty}M_{j,k}$. Since the function $t \mapsto t - \frac{1}{2}t^2$ is nondecreasing on $[0,1]$, by  taking $\limsup_{j \to \infty}$ on both sides of the recurrence inequality, we have $M \le M - \frac{1}{2}M^2$, implying $M=0$. Hence, $\lim_{j\to\infty}M_{j,k} = 0$.
	
	\paragraph{Distortion.}
	Recall that we chose $q$ so that $\alpha<\epsilon/4$. Choose sufficiently large $k$ such that $(1-\alpha)^{k+1} \le \alpha$. For the final winner $x_T$ and $c_p^* = c_p(v_0)$, let $E=E_{{v_0},k}$ be the event that the path from the root $v_0$ to $x_T$ contains no fatal errors and at most $k$ soft errors.
	Then in $E^c$, we have $S(x_T, c^*_p) \le \alpha$. Hence,
	$$\Exp[S(x_T,c_p^*)] \le \Pr(E) + \alpha \Pr(E^c) \le M_{d,k} + \alpha.$$
	By the preceding claim, choosing $d$ sufficiently large gives $\Exp[S(x_T,c_p^*)]<\epsilon/2$.
	Finally, for each realization of the pointers $p=(p_v)_v$, define voter-candidate distances by $d(i,c)= 1$ if $c \succsim_i c_p^*$ and $d(i,c)=3$ otherwise.
	Setting voter-voter distances and candidate-candidate distances to $2$ yields a valid metric.
	
	Under this metric, every candidate has social cost at least $n$, while $c_p^*$ has social cost exactly $n$.
	For a candidate $x$, $\sc(x)=n(3-2S(x,c_p^*))$ holds, and thus, 
	$$\Exp_p \left[\frac{ \Exp[\sc(x_T)| p]}{\sc(c^*_p)}\right] =3-2\Exp[S(x_T,c_p^*)]>3-\epsilon,$$
	where the inner expectation of the numerator is taken with respect to the algorithm's randomness.

	Therefore, there exists a realization $p^*$ of the pointers such that ${\Exp[\sc(x_T) | p=p^*]}/{\sc(c_{p^*}^*)}>3-\epsilon$.
	Since the pointers $p$ are sampled independently of the algorithm's randomness, $p^*$ and hence the fixed instance can be obtained through an oblivious adversary.
\end{proof}

Next, we consider adaptive adversaries.
Since the deterministic algorithm of \Cref{alg:online-weighted-uncovered} guarantees distortion of $2 + \sqrt{5} \approx 4.2361 < 5$, it is either that persistent randomized dictatorship is not optimal or the distortion $5$ analysis is not tight in the full information IC model.
It turns out that even under full information, the factor $5$ is tight, proving that persistent randomized dictatorship is not optimal under adaptive adversaries.

\begin{proposition}
	Under adaptive adversaries, \Cref{alg:persistent-random-dictatorship} cannot guarantee constant metric distortion below $5$ in the full information IC model.
\end{proposition}
\begin{proof}
		Fix a small $\epsilon > 0$. Consider the following instance for $n$ voters where $d(i,j)=2$ for all voters $i\ne j$. 
		The first candidate $c$ comes in with $d(i,c)=4$ for each voter $i$.
		Then for $i=1, \ldots, n$, candidate $c_i'$ comes in at round $i+1$, with distances $d(i,c_i')=3 - \epsilon$ and $d(j,c_i')=5-\epsilon$ for each $j \ne i$.
		Since $3 - \epsilon <4 < 5-\epsilon$, each voter $i\in N$ ranks $c_i' \succ_i c \succ_i c'_j$ for all $j \in [n] \setminus \{i\}$.
		Then after round $n+1$, the adversary can learn who the dictator is: $i^\star \in N$ with $x_{n+1}=c'_{i^\star}$. 
		Then at round $n+2$, the adversary introduces candidate $c^\star$ with distances $d(i^\star, c^\star)=3$ and $d(j, c^\star)=1$ for all $j \ne i^\star$. The remaining distances are given by the shortest path distances.
		
		At round $n+2$, since $3-\epsilon < 3$, we have $c_{i^\star}' \succ_{i^\star} c^\star$ and $x_{n+2}=c_{i^\star}'$. The social cost is
		$$\sc(c_{i^\star}')=(3-\epsilon) + (n-1)(5-\epsilon)=5n-2-n \epsilon,$$
		whereas
		$$\sc(c^\star)=3 + (n-1)\cdot 1 =n+2.$$
		Hence, the metric distortion is $$\frac{5n-2-n\epsilon}{n+2} \to \frac{5n-2}{n+2},$$ as $\epsilon \to 0$. Let $n \to \infty$.
\end{proof}

This leaves the optimal distortion bound for adaptive adversaries to be in the interval $3 \le \Delta^* \le 2 + \sqrt{5}$. The lower bound comes from the lower bound from an oblivious adversary, and the upper bound comes from a deterministic algorithm. We leave it as an open problem to close the gap. Since the lower bound did not utilize adaptivity of the adversary and the upper bound did not utilize randomness, we conjecture the optimal distortion to lie strictly between the two numbers $\Delta^* \in (3, 2 + \sqrt{5})$.

\section{Conclusion}
We studied metric distortion under online candidate arrivals. This provided the online algorithmic perspective to the theory of metric distortion. Based on two dimensions of information and action space, we characterized rules that yield optimal distortion in both deterministic and randomized settings, except for randomized voting rules against adaptive adversaries in the full information IC model. Closing the gap remains open. 
As future work, considering the utilitarian distortion under the two dimensions, along with two adversaries, will be interesting.

\section*{Acknowledgments}
Shinsaku Sakaue was supported by the JST BOOST Program (Japan), Grant Number JPMJBY24D1.

\bibliography{online_distortion}

@inproceedings{procaccia2006distortion,
  author    = {Ariel D. Procaccia and Jeffrey S. Rosenschein},
  title     = {The Distortion of Cardinal Preferences in Voting},
  booktitle = {Proceedings of the 10th International Workshop on Cooperative Information Agents (CIA)},
  series    = {Lecture Notes in Computer Science},
  volume    = {4149},
  pages     = {317--331},
  publisher = {Springer},
  year      = {2006},
  doi       = {10.1007/11839354_23}
}

@inproceedings{munagala2019weighted,
  author    = {Kamesh Munagala and Kangning Wang},
  title     = {Improved Metric Distortion for Deterministic Social Choice Rules},
  booktitle = {Proceedings of the 2019 ACM Conference on Economics and Computation (EC)},
  pages     = {245--262},
  publisher = {ACM},
  year      = {2019},
  doi       = {10.1145/3328526.3329550},
  eprint    = {1905.01401},
  archivePrefix = {arXiv},
  primaryClass  = {cs.GT}
}

@inproceedings{gkatzelis2020optimal,
	author        = {Gkatzelis, Vasilis and Halpern, Daniel and Shah, Nisarg},
	title         = {Resolving the Optimal Metric Distortion Conjecture},
	booktitle     = {Proceedings of the 61st {IEEE} Annual Symposium on Foundations of Computer Science ({FOCS})},
	year          = {2020},
	publisher     = {IEEE},
	pages         = {1427--1438},
	doi           = {10.1109/FOCS46700.2020.00134},
	eprint        = {2004.07447},
	archivePrefix = {arXiv},
	primaryClass  = {cs.GT}
}

@inproceedings{kizilkaya2022plurality,
  author    = {Fatih Erdem Kizilkaya and David Kempe},
  title     = {Plurality Veto: A Simple Voting Rule Achieving Optimal Metric Distortion},
  booktitle = {Proceedings of the Thirty-First International Joint Conference on Artificial Intelligence (IJCAI)},
  pages     = {349--355},
  year      = {2022},
  doi       = {10.24963/ijcai.2022/50},
  eprint    = {2206.07098},
  archivePrefix = {arXiv},
  primaryClass  = {cs.GT}
}

@article{anshelevich2016randomized,
	author  = {Elliot Anshelevich and John Postl},
	title   = {Randomized Social Choice Functions under Metric Preferences},
	journal = {Journal of Artificial Intelligence Research},
	volume  = {58},
	pages   = {797--827},
	year    = {2017},
	doi     = {10.1613/jair.5340}
}

@inproceedings{charikar2022randomized,
	author        = {Charikar, Moses and Ramakrishnan, Prasanna},
	title         = {Metric Distortion Bounds for Randomized Social Choice},
	booktitle     = {Proceedings of the 2022 Annual {ACM-SIAM} Symposium on Discrete Algorithms ({SODA})},
	year          = {2022},
	publisher     = {SIAM},
	pages         = {2986--3004},
	doi           = {10.1137/1.9781611977073.116},
	eprint        = {2111.03694},
	archivePrefix = {arXiv},
	primaryClass  = {cs.GT}
}

@article{charikar2024barrier,
  author  = {Moses Charikar and Prasanna Ramakrishnan and Kangning Wang and Hongxun Wu},
  title   = {Breaking the Metric Voting Distortion Barrier},
  journal = {Journal of the ACM},
  volume  = {71},
  number  = {6},
  pages   = {42:1--42:33},
  year    = {2024},
  doi     = {10.1145/3689625},
  eprint  = {2306.17838},
  archivePrefix = {arXiv},
  primaryClass  = {cs.GT}
}

@inproceedings{Kempe2020LPDuality,
	author    = {Kempe, David},
	title     = {An Analysis Framework for Metric Voting based on {LP} Duality},
	booktitle = {Proceedings of the {AAAI} Conference on Artificial Intelligence},
	year      = {2020},
	volume    = {34},
	number    = {2},
	pages     = {2079--2086},
	doi       = {10.1609/aaai.v34i02.5581},
	url       = {https://doi.org/10.1609/aaai.v34i02.5581}
}

@misc{ghasemi2026distortion,
	title         = {Distortion in Online Voting},
	author        = {Ghasemi, MohammadHossein and Seddighin, Masoud and Seddighin, Saeed},
	year          = {2026},
	eprint        = {2611.25709},
	archivePrefix = {arXiv},
	primaryClass  = {cs.GT},
	url           = {https://arxiv.org/abs/2611.25709}
}

@article{caragiannis2011voting,
	author  = {Ioannis Caragiannis and Ariel D. Procaccia},
	title   = {Voting Almost Maximizes Social Welfare Despite Limited Communication},
	journal = {Artificial Intelligence},
	volume  = {175},
	number  = {9--10},
	pages   = {1655--1671},
	year    = {2011},
	doi     = {10.1016/j.artint.2011.03.005}
}

@article{boutilier2015optimal,
	author  = {Craig Boutilier and Ioannis Caragiannis and Simi Haber and Tyler Lu and Ariel D. Procaccia and Or Sheffet},
	title   = {Optimal Social Choice Functions: A Utilitarian View},
	journal = {Artificial Intelligence},
	volume  = {227},
	pages   = {190--213},
	year    = {2015},
	doi     = {10.1016/j.artint.2015.06.003}
}

@article{ebadian2022optimized,
	author  = {Soroush Ebadian and Anson Kahng and Dominik Peters and Nisarg Shah},
	title   = {Optimized Distortion and Proportional Fairness in Voting},
	journal = {ACM Transactions on Economics and Computation},
	volume  = {12},
	number  = {1},
	pages   = {3:1--3:39},
	year    = {2024},
	doi     = {10.1145/3640760}
}

@inproceedings{anshelevich2021survey,
	author    = {Elliot Anshelevich and Aris Filos-Ratsikas and Nisarg Shah and Alexandros A. Voudouris},
	title     = {Distortion in Social Choice Problems: The First 15 Years and Beyond},
	booktitle = {Proceedings of the Thirtieth International Joint Conference on Artificial Intelligence (IJCAI)},
	pages     = {4294--4301},
	year      = {2021},
	doi       = {10.24963/ijcai.2021/589},
	eprint    = {2103.00911},
	archivePrefix = {arXiv}
}

@article{anshelevich2015metric,
	author  = {Elliot Anshelevich and Onkar Bhardwaj and Edith Elkind and John Postl and Piotr Skowron},
	title   = {Approximating Optimal Social Choice under Metric Preferences},
	journal = {Artificial Intelligence},
	volume  = {264},
	pages   = {27--51},
	year    = {2018},
	doi     = {10.1016/j.artint.2018.07.006}
}

@misc{cai2026stable,
	author        = {Cai, Ziyi and Frank, Fabian and Hastings, Jabari and Ramakrishnan, Prasanna and Wang, Kangning and Ye, Qilin},
	title         = {Stable Voting Rules on the Edge of Optimal Metric Distortion},
	year          = {2026},
	eprint        = {2609.08259},
	archivePrefix = {arXiv},
	primaryClass  = {cs.GT},
	url           = {https://arxiv.org/abs/2609.08259v2},
	note          = {Version 2, November 3, 2026}
}

@inproceedings{charikar2025tournament,
	author    = {Moses Charikar and Prasanna Ramakrishnan and Zihan Tan and Kangning Wang},
	title     = {Metric Distortion for Tournament Voting and Beyond},
	booktitle = {Proceedings of the 26th ACM Conference on Economics and Computation (EC)},
	pages     = {790--818},
	publisher = {ACM},
	year      = {2025},
	doi       = {10.1145/3736252.3742629},
	eprint    = {2505.13630},
	archivePrefix = {arXiv}
}

@misc{lin2026weighted,
	author        = {Lin, Jianan and Wang, Chenhao},
	title         = {Improved Metric Distortion Bounds for Deterministic Weighted-Tournament Voting Rules},
	year          = {2026},
	eprint        = {2608.15247},
	archivePrefix = {arXiv},
	primaryClass  = {cs.GT},
	url           = {https://arxiv.org/abs/2608.15247v2},
	note          = {Version 2, November 11, 2026}
}

@inproceedings{pierczynski2019approval,
	author    = {Grzegorz Pierczy{\'n}ski and Piotr Skowron},
	title     = {Approval-Based Elections and Distortion of Voting Rules},
	booktitle = {Proceedings of the Twenty-Eighth International Joint Conference on Artificial Intelligence (IJCAI)},
	pages     = {543--549},
	year      = {2019},
	doi       = {10.24963/ijcai.2019/77},
	eprint    = {1901.06709},
	archivePrefix = {arXiv}
}

@article{anshelevich2024threshold,
	author  = {Elliot Anshelevich and Aris Filos-Ratsikas and Christopher Jerrett and Alexandros A. Voudouris},
	title   = {Improved Metric Distortion via Threshold Approvals},
	journal = {Artificial Intelligence},
	volume  = {341},
	pages   = {104295},
	year    = {2025},
	doi     = {10.1016/j.artint.2025.104295}
}

@misc{frank2026approval,
	author        = {Fabian Frank and Jannik Peters},
	title         = {A Constant Metric Distortion Protocol for Approval Voting Given Plurality Polls},
	year          = {2026},
	eprint        = {2608.28340},
	archivePrefix = {arXiv},
	primaryClass  = {cs.GT},
	url           = {https://arxiv.org/abs/2608.28340}
}

@inproceedings{oren2014online,
	author    = {Joel Oren and Brendan Lucier},
	title     = {Online (Budgeted) Social Choice},
	booktitle = {Proceedings of the Twenty-Eighth AAAI Conference on Artificial Intelligence (AAAI)},
	pages     = {1456--1462},
	publisher = {AAAI Press},
	year      = {2014},
	doi       = {10.1609/aaai.v28i1.8891}
}

@inproceedings{dong2025maintaining,
	author    = {Chris Dong and Jannik Peters},
	title     = {Maintaining Proportional Committees with Dynamic Candidate Sets},
	booktitle = {Proceedings of the 42nd International Conference
	on Machine Learning (ICML)},
	series    = {Proceedings of Machine Learning Research},
	volume    = {267},
	pages     = {14152--14168},
	publisher = {PMLR},
	year      = {2025},
	url       = {https://proceedings.mlr.press/v267/dong25j.html}
}

@article{black1948rationale,
	author  = {Duncan Black},
	title   = {On the Rationale of Group Decision-making},
	journal = {Journal of Political Economy},
	volume  = {56},
	number  = {1},
	pages   = {23--34},
	year    = {1948},
	doi     = {10.1086/256633}
}

@article{riker1958paradox,
	author  = {William H. Riker},
	title   = {The Paradox of Voting and Congressional Rules for Voting on Amendments},
	journal = {American Political Science Review},
	volume  = {52},
	number  = {2},
	pages   = {349--366},
	year    = {1958},
	doi     = {10.2307/1952321}
}

@book{farquharson1969theory,
	author    = {Robin Farquharson},
	title     = {Theory of Voting},
	publisher = {Blackwell},
	year      = {1969}
}

@article{mckelvey1978multistage,
	author  = {Richard D. McKelvey and Richard G. Niemi},
	title   = {A Multistage Game Representation of Sophisticated Voting for Binary Procedures},
	journal = {Journal of Economic Theory},
	volume  = {18},
	number  = {1},
	pages   = {1--22},
	year    = {1978},
	doi     = {10.1016/0022-0531(78)90039-X}
}

@article{banks1985sophisticated,
	author  = {Jeffrey S. Banks},
	title   = {Sophisticated Voting Outcomes and Agenda Control},
	journal = {Social Choice and Welfare},
	volume  = {1},
	number  = {4},
	pages   = {295--306},
	year    = {1985},
	doi     = {10.1007/BF00649265}
}

@article{bredereck2017parliamentary,
	author  = {Robert Bredereck and Jiehua Chen and Rolf Niedermeier and Toby Walsh},
	title   = {Parliamentary Voting Procedures: Agenda Control, Manipulation, and Uncertainty},
	journal = {Journal of Artificial Intelligence Research},
	volume  = {59},
	pages   = {133--173},
	year    = {2017},
	doi     = {10.1613/jair.5407}
}

@article{miller1977graph,
	author  = {Nicholas R. Miller},
	title   = {Graph-Theoretical Approaches to the Theory of Voting},
	journal = {American Journal of Political Science},
	volume  = {21},
	number  = {4},
	pages   = {769--803},
	year    = {1977},
	doi     = {10.2307/2110736}
}

\clearpage

\appendix
\subsection*{AI Usage Disclosure}
We have used AI models (ChatGPT 6.2 Astra, 6.1 Sol, 6 Astra, 5.6 Sol, 5.5, and 5.4) in the following ways. First,
we have used them to search for related work on the topic. Second, we have used them to enhance the wording and exposition of the authors’ drafts. Third, we used them as an additional check of the proofs written by the authors. Finally, we have used them to derive bounds (mainly lower bounds) for our models. Starting from weaker bounds supplied by the authors, the models helped strengthen some bounds to tight results. Although some bounds were derived by AI, we have independently checked all final arguments and have rewritten the proofs extensively to make the proofs accessible to human readers. The authors take responsibility for them. The model of online candidate arrivals with two dimensions of information and action space, together with two different adversaries, was proposed by the authors.

\section{Appendix: Full Candidate Model with Partial Information}\label{sec:ic-fi}
We now consider the partial information full candidate model. As noted in \Cref{sec:partial-info}, the optimal bounds for deterministic algorithms and randomized algorithms under adaptive adversaries naturally extend without changing the proof.

To extend the lower bound of $3$ for randomized algorithms against oblivious adversaries to the full candidate model, we first establish the following lemma.
\begin{lemma}
	Suppose a randomized voting rule has constant distortion.
	If all voters prefer $c_t$ to $x_{t-1}$, then for each $s \ge t$, $x_s \in \{c_t, c_{t+1}, \ldots, c_T\}$ almost surely in the partial information full candidate model.
\end{lemma}
\begin{proof}
	Assume to the contrary that for some instance, round $t$, and $s \ge t$, all voters prefer $c_t$ to $x_{t-1}$ but $x_s \notin \{c_t, c_{t+1}, \ldots, c_T\}$ with positive probability $p > 0$. 
	Let $s$ be the first $s \ge t$ to satisfy such a condition.
	Since there are finitely many possible observations and decision histories up until round $s$, there exists a history $h$ occurring with probability $q > 0$ and satisfying the conditions.

	Let $M > 1$. Consider another instance on a line where all voters are at $0$, candidates $c_1, \ldots, c_{t-1}$ are at $M$, and all other candidates $c_t, c_{t+1}, \ldots, c_T$ are at $1$. 
	Then all voters prefer $c_t$ to $x_{t-1}$. Break ties within $C_{t-1}=\{c_1, \ldots, c_{t-1}\}$ to match the preference profile of the original instance. Hence, up until round $t$, the algorithm cannot distinguish the two instances.
	If $s > t$, then in history $h$, $c_t$ is selected. Hence for rounds $t+1, \ldots, s$, we only compare candidates in $\{c_t, c_{t+1}, \ldots, c_s\}$. Hence, we can break ties within $\{c_t, c_{t+1}, \ldots, c_s\}$ to match the preference profile of the original instance. Hence, the algorithm cannot distinguish the two instances when it simulates the history $h$ with positive probability $q$.

	Then at round $s$, the selected old candidate in $\{c_1, \ldots, c_{t-1}\}$ has social cost $nM$, whereas an optimal candidate has $n$. Thus, expected distortion is at least $qM$, which grows to infinity as $M \to \infty$, a contradiction.
\end{proof}
This allows us to transfer the lower bound of $3$.

\begin{corollary}
	Under oblivious adversaries, no randomized voting rule can guarantee a constant metric distortion below $3$ in the partial information full candidate model. 
\end{corollary}
\begin{proof}
	We modify the construction in the proof of \Cref{thm:pi-ic-obliv}.
	Previously, for each candidate $c \ne c_1$, we had an associated set $B(c)$. Instead, we duplicate each candidate and let two consecutive candidates share the same associated set.
	 
	Fix $n \ge 2$ and $D < 3-\frac{2}{n}$, and suppose a randomized voting rule guarantees distortion at most $D$.
	Consider the same metric and sampling distribution as in the previous proof. Set $B(c_1)=N$. For each $r \ge 1$, independently sample $S_r \subseteq N$ according to
	$$\Pr(S_r=S)=\frac{\lambda^{-|S|}}{Z},$$
	where $S \subseteq N$ and $Z=(1+\lambda^{-1})^n$.
	Set $B(c_{2r})=B(c_{2r+1})=S_r$. All sets are sampled in advance. Recall the potential function $\phi(k)=\frac{Z}{\lambda-1}\lambda^k$ and $\eta=\Exp[\phi(|S_r|)]=\frac{2^n}{\lambda-1} > 0$.
	
	We modify the potential to account for the possibility of returning to previous candidates, which can happen in algorithms working under the full candidate space. Whenever the algorithm selects $c_{2r}$, all voters prefer $c_{2r+1}$ to $c_{2r}$, since they have the same associated set $S_r$ and ties favor later arrivals. By the preceding lemma, all candidates preceding $c_{2r+1}$ become permanently unavailable. Call this event a \emph{reset}.
	
	Set $M_0=n$, and for each $r\ge1$, define
	$$
	M_r=
	\begin{cases}
		|S_r|,&\text{if }x_{2r}=c_{2r},\\
		\max\{M_{r-1},|S_r|\},&\text{otherwise}.
	\end{cases}
	$$
	The quantity $M_r$ bounds the sizes of the associated sets of all candidates that the algorithm may still select among those that have arrived. Indeed, after a reset, only $c_{2r+1}$ remains available among the candidates seen so far. Otherwise, the previous bound is preserved, and the two new candidates both have associated-set size $|S_r|$. We track the potential $\phi(M_r)$. (Compare to how we tracked $\phi(|B(x_t)|)$ before.)
	
	Fix $r \ge 1$, and let $b=|B(x_{2r-1})|$. By the preceding observation, $b\le M_{r-1}$. For each $i\in N$, let $X_i$ indicate that the algorithm selects a candidate other than the challenger $c_{2r}$ when exactly voter $i$ prefers the incumbent $x_{2r-1}$ at round $2r$. As in the previous proof, we then have
	$$\Exp[X_i]\le\frac{D-1}{2(n-1)}.$$
	
	We analyze the expected change in $\phi(M_r)$. Condition on the previously sampled sets and randomness of the algorithm so that $B(x_{2r-1})$, $b$, $M_{r-1}$, and $X_i$ are determined. For each $i\in B(x_{2r-1})$, consider the event $E_i=\{S_r= B(x_{2r-1}) \setminus\{i\}\}$, and let $E=\bigcup_{i\in B(x_{2r-1})}E_i$. These events are disjoint.
	
	On $E_i$, exactly voter $i$ prefers $x_{2r-1}$ to $c_{2r}$. If $X_i=0$, the algorithm selects $c_{2r}$, causing a reset, so $M_r=b-1$. Otherwise, $X_i=1$, and $M_r=M_{r-1}$, since $b-1\le M_{r-1}$. Therefore, on $E_i$,
	$$ \phi(M_r)-\phi(M_{r-1}) \le (1-X_i)(\phi(b-1)-\phi(b)).$$
	Since $S_r$ is sampled independently of the previous sets and the algorithm's randomness, under the conditioning, we have $\Pr(E_i)=\frac{\lambda^{-(b-1)}}{Z}$.
	This yields
	$$\Pr(E_i)(\phi(b)-\phi(b-1)) =\frac{\lambda^{-(b-1)}}{Z}(\phi(b)-\phi(b-1))=1,$$
	and hence the conditional expected contribution of $E_i$ is at most $-(1-X_i)$.
	
	Moreover, by the definition of $M_r$, we always have $M_r\le\max\{M_{r-1},|S_r|\}$, and hence, since $\phi$ is nonnegative and increasing,
	$$\phi(M_r)-\phi(M_{r-1})\le\phi(|S_r|).$$
	Thus, the remaining events contribute at most $\eta$ in expectation. Combining these bounds and removing the conditioning, we obtain
	\begin{align*}
		\Exp[\phi(M_r)-\phi(M_{r-1})]
		&\le -\Exp[b]+\Exp\left[\sum_{i\in B(x_{2r-1})}X_i\right]+\eta\\
		&\le -\Exp[b]+\sum_{i \in N}\Exp\left[X_i\right]+\eta\\
		&\le -\Exp[b]+\beta+\eta,
	\end{align*}
	where $\beta=\frac{n(D-1)}{2(n-1)}<1$, as in the previous proof.	
	
	The remaining arguments proceed analogously.
	Since $c_1$ remains optimal with social cost $n$, we have $\Exp[b]\ge\frac{n(3-D)}2>1$.
	Taking sufficiently large $\lambda$ so that $\delta=1-\beta-\eta>0$, we obtain $\Exp[\phi(M_r)-\phi(M_{r-1})]\le-\delta$.
	Summing over $r=1,\ldots,R$ yields $\Exp[\phi(M_R)]\le\phi(n)-R\delta$, and taking sufficiently large $R$ makes the right-hand side negative, contradicting the nonnegativity of $\phi$.
\end{proof}

\end{document}